\documentclass[12pt]{article}

\usepackage[margin=1in]{geometry}
\usepackage{setspace}
\usepackage{amsmath,amssymb}
\usepackage{booktabs,threeparttable,longtable,array,tabularx}
\usepackage{graphicx}
\graphicspath{{supporting_files/}}
\usepackage{caption}
\usepackage{subcaption}
\usepackage{authblk}
\usepackage[sort,longnamesfirst]{natbib}
\bibpunct{(}{)}{;}{a}{}{,}
\usepackage{hyperref}
\usepackage{pdflscape}
\usepackage{float}
\usepackage{enumitem}
\usepackage{microtype}

\hypersetup{colorlinks=true, linkcolor=black, citecolor=black, urlcolor=blue,
  pdftitle={Using AI Predictions to Augment Rather Than Replace Surveys},
  pdfauthor={Lai Wei and Yifei Zhu},
  pdfkeywords={prediction-powered inference, survey methodology, large language models, machine learning, statistical efficiency, auxiliary data}}
\newcommand{\placehere}[1]{\par\begin{center}[#1 about here]\end{center}\par}

\title{Using AI Predictions to Augment Rather Than Replace Surveys}
\author[1]{Lai Wei}
\author[2]{Yifei Zhu}
\affil[1]{Department of Sociology, The University of Hong Kong\\
  \href{mailto:laiw@hku.hk}{laiw@hku.hk}}
\affil[2]{Department of Politics and Public Administration,\\
  The University of Hong Kong\\
  \href{mailto:frankyifei@connect.hku.hk}{frankyifei@connect.hku.hk}}

\date{}

\begin{document}
\maketitle


\noindent\textbf{Transparency and Reproducibility Statement.}
The replication repository for this study is hosted on the Open Science Framework (\href{https://osf.io/fkvst/overview?view_only=9c8f5e11d7ca44a4834985580102d6b9}{OSF}). 

\noindent\textbf{Acknowledgements.}
This research was supported by the Research Grants Council of the Hong Kong Special Administrative Region, China, through its Early Career Scheme (Project No.~27623826, ``Powering Survey Research with Census Data and Predictions from Machine Learning and LLM'').

\clearpage
\begin{abstract}
LLMs can predict human survey responses with nontrivial accuracy. What this means for the future of social surveys remains hotly contested. Optimists anticipate a coming era of synthetic respondents; skeptics see a fragile and overhyped substitute. This article advocates a position: AI predictions cannot replace surveys, but they can augment them. Using prediction-powered inference (PPI), a statistical framework rooted in survey methodology and known in sociology as ``mixed subjects design'', the study tests whether AI predictions can improve the statistical efficiency of social surveys when combined with census-scale auxiliary information. Applying the framework to 83 social outcomes in the General Social Survey (GSS) and National Health Interview Survey (NHIS), with American Community Survey (ACS) being the auxiliary file, we find that supervised predictors yield effective-sample-size gains of about 7--12 percent when estimating the mean, while LLM gains are smaller but positive. 

\end{abstract}

\section{Introduction}

AI is no longer a distant future; it is already part of the infrastructure of social science. Across fields, predictive systems now support tasks ranging from classification to simulation. Sociology has entered this moment as well, as generative models increasingly shape debates about measurement and routine research practice \citep{bail2024,alvero2026}. 

A separate trend sociologists are facing is the increasing difficulty of fielding social surveys. Probability surveys remain an indispensable source of information for understanding society, but the classical operating model of large samples, high response rates, and stable fieldwork costs has become harder to sustain \citep{groves2006,groves2008,nationalresearchcouncil2013,dutwin2021}.

When these two trends meet, we see an emerging interest in the use of AI for social surveys. Large language models can be prompted with demographic profiles and then requested to answer common social survey questions, and they often produce responses that resemble human answers in nontrivial ways \citep{argyle2023,santurkar2023,bisbee2024,kimlee2023,rosenbusch2023,xie2026}. Supervised learning trained on earlier survey waves can also predict many survey outcomes from demographic and socioeconomic covariates \citep{salganik2020,zheng2025,rahal2024}. These results make it tempting to imagine a future in which surveys become smaller, with predictive models filling in what humans would have said.

That temptation requires caution. AI-generated survey responses can contain arbitrary and hard-to-audit bias. They can be too centered or under-dispersed, muting the heterogeneity that surveys are designed to measure. They can also be too discriminative, reproducing stereotypes or group-specific distortions in the name of prediction \citep{santurkar2023,bisbee2024,kozlowski2025,westwood2025,xie2026}. 

Treating AI predictions as replacements for survey responses gives models more authority than their bias can support. Discarding those predictions altogether leaves useful information unused. We advocate a middle path, using AI predictions to augment actual surveys. A prediction can be wrong in level or distribution and still be useful if it is correlated with the true outcome. That correlation lets the model summarize the part of the outcome that is predictable from observed covariates in a large auxiliary file. The probability survey then measures prediction error by comparing predicted and observed responses among sampled humans. The resulting estimator uses AI for the predictable component and uses the survey to correct what AI misses, preserving design-based validity while reducing sampling variability.

Prediction-powered inference (PPI) formalizes this augmentation strategy \citep{angelopoulos2023,angelopoulos2024,zrnic2024}. Its central operation is outcome prediction followed by correction. Predictions on a large unlabeled file are combined with a correction estimated from observed outcomes in a smaller labeled sample. In our application, we also incorporate survey weights to account for the sampling designs, whereas prediction-powered augmentation uses additional information to improve precision. For mean estimation, this correction follows the logic of the classical difference estimator, linking PPI to the model-assisted survey tradition \citep{cassel1976,deville1992,sarnswenssonwretman1992,breidt2017,mozer2026}.

PPI is especially useful when collecting additional observed outcomes is expensive but a much larger unlabeled file from the same population distribution is available. The objective is to improve precision for the same inferential target, not to establish representativeness or justify extrapolation to a different population. Gains depend on whether predictions sufficiently reduce the variation that must be estimated from the labeled sample; a large unlabeled file alone does not guarantee improvement. Social surveys offer an important application because additional interviews are costly, while files such as the American Community Survey (ACS), harmonized through IPUMS, provide compatible covariates for many more individuals without measuring the focal survey outcomes.

We test how much PPI can help with routine quantitative tasks in sociology, including estimating population means and estimating regression coefficients. We apply the framework to a range of opinion outcomes in the General Social Survey (GSS) and health outcomes in the National Health Interview Survey (NHIS), with ACS as the auxiliary file. We compare historical supervised learning, cross-fitted supervised learning, and LLM respondent simulation, and we evaluate both predictive performance and downstream inferential gain. For mean estimation, supervised predictors yield effective-sample-size gains of about 7 to 12 percent, while LLM gains are smaller. In this benchmark, gains are generally larger for health outcomes than opinion outcomes and for means than regression coefficients. These differences matter for how researchers use PPI: a predictor that improves estimation of an outcome mean may provide little benefit for an adjusted demographic association. Predictors should therefore be evaluated for the intended estimand, rather than by outcome-level predictive accuracy alone.

In summary, this article illustrates a design principle for AI-era survey research, which uses cheap AI predictions to augment expensive human responses. The analyses show how predictions can be incorporated into probability survey inference without treating model outputs as respondents. The empirical results establish a conservative baseline using public surveys, ACS covariates, and contemporary prediction tools. Better auxiliary data and stronger prediction systems may increase those gains, while weakly predicted outcomes and groups identify where human measurement remains indispensable. We invite survey methodologists, sociologists, and machine-learning researchers to test this design in new surveys and build richer auxiliary frames. We also invite researchers to release PPI-ready prediction artifacts and evaluate predictors by the precision they add after survey data correct prediction error. These exercises will help improve the statistical efficiency of social surveys without conducting additional interviews, thus reducing costs for survey organizers and improving power for data users.

\section{Related Literature}

Three literatures define the problem this article addresses. Survey methodology shows how auxiliary information can improve precision when probability sampling remains the source of validity. The predictive turn in social science shows that many social outcomes contain measurable but uneven signal. Research on LLM-generated respondents shows that AI predictions are cheap and flexible, but also that uncorrected synthetic responses cannot support population inference. 

\subsection{Model-assisted and prediction-powered inference}

Survey methodology has long allowed models to improve precision without giving them final authority over the estimand. Poststratification and ratio adjustment align survey estimates with known population totals. Other model-assisted estimators use working models, adjusted weights, or modeled cells to exploit auxiliary covariates \citep{holt1979,cassel1976,deville1992,sarnswenssonwretman1992,wu2001,breidt2017,little1993,gelman1997,park2004,laxphillips2009,ghitzagelman2013,mozer2026}. Their implementations differ, but the inferential discipline is the same. Model structure can lower variance only because the probability sample still defines the population target and supplies a correction for model error.

Prediction-powered inference gives this correction logic a modern form. In the canonical unweighted PPI setting, a small labeled sample contains gold-standard outcomes and a larger unlabeled file contains covariates, with both samples drawn from a common distribution. An external predictor assigns provisional outcomes to both files \citep{angelopoulos2023}. The estimator does not analyze predicted outcomes as if they were observations. It uses predictions on the large file to estimate the predictable component, then uses the labeled sample to estimate and correct prediction error. This construction does not require reweighting the labeled sample; survey weights enter our application separately to account for its sampling designs.

A related design has entered sociology under the term ``mixed subjects design'', which applies PPI to combine human responses with LLM predictions \citep{broska2025}. That work establishes a way to use synthetic responses without treating them as a substitute for human data. What remains to be demonstrated is whether the residual-correction logic can improve probability-sample inference for social surveys. This article helps establish the applicability of mixed subjects design in large and public social survey settings, in combination with population censuses.

PPI is one approach within the broader literature on inference with predicted data. \citet{salerno2025} review how these methods address bias and uncertainty and connect them to survey sampling, missing-data analysis, measurement error, and semi-supervised inference. Semi-supervised inference studies how unlabeled covariates can improve estimation of means and regression coefficients \citep{chakrabortty2018,zhang2019,azriel2022}. Post-prediction and surrogate-label methods show how downstream inference can be biased when predicted outcomes are used without correction \citep{wang2020,egami2023}. Orthogonalized estimators similarly separate flexible nuisance prediction from valid inference for target parameters \citep{chernozhukov2018}. Recent PPI extensions improve rectification and allocation while extending the framework to harder outcomes, missing covariates, nonuniform sampling, and survey-like LLM allocation problems \citep{angelopoulos2024,zrnic2024,zrnic2024active,demirel2024,fan2024,kluger2025,li2025ppi,ye2026}. We use PPI's residual-correction formulation to evaluate augmentation of probability surveys with a large auxiliary file.

\subsection{Predictability as a survey-design quantity}

Auxiliary predictions improve precision when they reduce the variation the labeled survey must estimate. A recent literature in sociology asks how much social outcomes can be predicted at all. The predictive turn in social science argues that out-of-sample prediction is a useful complement to explanation because it exposes whether models generalize beyond the data on which they were fit \citep{yarkoni2017,hofman2021,verhagen2022,rahal2024}. The Fragile Families Challenge is the landmark benchmark in this literature. In a common-task competition using a rich longitudinal cohort, 160 teams predicted six age-15 outcomes from thousands of earlier-life covariates \citep{salganik2019,salganik2020}. Even the best models achieved only modest holdout performance. The out-of-sample R-square is roughly .23 for material hardship and .19 for GPA, with lower values for the remaining outcomes. Subsequent papers showed that intensive feature engineering and flexible algorithms helped, but often only modestly relative to simpler baselines \citep{rigobon2019,stanescu2019,mckay2019,carnegie2019}.

Those results have usually been interpreted as evidence about social complexity rather than as inputs to survey design. Qualitative follow-up to the Fragile Families Challenge emphasizes contingency, timing, and local context beyond structured covariates \citep{lundberg2024,small2024}. \citet{yan2025} distinguish irreducible uncertainty from prediction errors that could shrink with better measurement, while studies across several domains show that predictability varies with the outcome and research setting \citep{kuikka2024,zheng2025,domingue2025,badolato2024}. These studies take predictability as a quantity that contains theoretical meaning by itself and link it to concepts like social rigidity or contingency. 

This article uses predictability as a survey-design quantity. Earlier scholars reported how predictable a range of life outcomes are, and what predictability tells us about society. We ask how these predictions can improve the efficiency of our usual research tasks. A modest $R^2$ may be unpersuasive for individual forecasting but still valuable for survey estimation when predictions can be generated at auxiliary scale. 

\subsection{LLMs as external predictors for survey augmentation}

Survey augmentation requires predictions that are external to the outcomes being rectified. Supervised machine learning can supply such predictions when repeated survey modules exist, but many modules are new or redesigned across time. LLMs relax this constraint because a respondent profile can be paired with the question and response scale to create a prediction task without first collecting a large outcome-specific training set. This makes LLMs attractive for augmented surveys, especially when researchers need provisional predictions before enough human labels exist for conventional supervised learning.

A growing literature asks whether LLMs can simulate survey respondents or experiment participants \citep{argyle2023,aher2023,horton2023,rosenbusch2023}. This literature sheds initial light on the prospect of using LLMs for survey augmentation. LLMs can recover nontrivial structure in human responses across many tasks \citep{kimlee2023,cao2025,zhao2025,yang2024}. These findings imply that LLM outputs may contain enough signal to be useful after correction.

The same literature explains why correction is mandatory. LLM-generated responses can be under-dispersed and demographically misaligned. They can also shift over time and across prompts \citep{santurkar2023,durmus2023,dominguez2024,bisbee2024,boelaert2025,kozlowski2025}. A recent high-profile study shows that LLM-generated social science data can appear realistic while failing to reproduce the distributional, associational, and sequential structure needed for social inference \citep{xie2026}. Related work warns that AI-generated content can homogenize social data and threaten survey integrity \citep{xiehomogenization2026,westwood2025,zhang2025}. Methodological critiques therefore separate exploratory uses such as piloting and design checks from confirmatory population inference, which requires validated and transparent calibration \citep{rossi2024,abdurahman2025,hullman2026,krsteski2025}. This failure pattern makes augmentation preferable to replacement. A probability sample can estimate the gap between predicted and observed responses, but only if human responses remain the source of validity.

\section{Method}

\begin{figure}[p]
\centering
\includegraphics[width=0.95\textwidth]{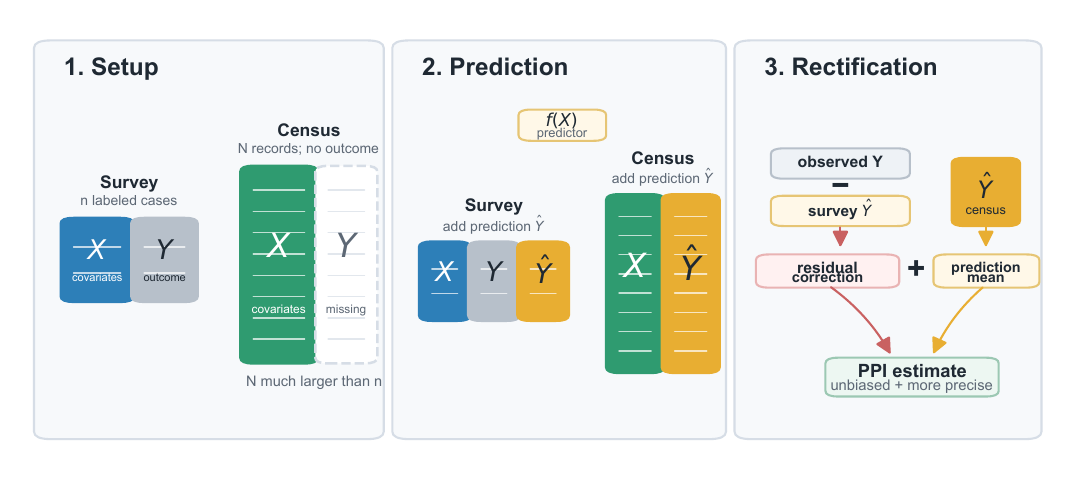}
\caption{Basic workflow of prediction-powered survey augmentation. The survey observes both covariates $X$ and outcomes $Y$ for a relatively small labeled sample. The census-scale auxiliary file is much larger but contains covariates only. A predictor $f(X)$ supplies predicted outcomes $\hat{Y}$ for both sources. PPI combines the census prediction mean with the survey-estimated residual correction, producing an unbiased estimate that can be more precise when predictions contain useful signal.}
\label{fig:ppi-illustration}
\end{figure}

\placehere{Figure~\ref{fig:ppi-illustration}}

Figure \ref{fig:ppi-illustration} gives the basic workflow of prediction-powered survey augmentation. The survey has both covariates and outcomes of interest. Apart from the survey, we also have an auxiliary file that is much larger than the survey, referred to as the census. The census need not literally include everyone in the population. The census has only covariates but not outcomes. We also have access to an external predictor that takes covariates as input and predicts outcomes.  We predict outcomes for both survey respondents and census respondents, for whom outcomes are uncollected. The final estimator uses both pieces: a prediction mean over the large auxiliary file and a residual correction from the labeled survey. We first present this idea in the simplest unweighted mean case, because it makes the correction and efficiency mechanism transparent, and then extend the same logic to weighted surveys and regression coefficients.

\subsection{Prediction-powered estimation}

Suppose the target is the population mean $\mu=E(Y)$. A labeled survey sample of size $n$ contains both covariates and outcomes, $(X_i,Y_i)$. A much larger auxiliary file of size $N$ contains only covariates, $\tilde X_j$. For this unweighted example, assume the survey and auxiliary samples are independent random samples from the same population distribution. Let $f(\cdot)$ be an external predictor of the outcome from covariates. The ordinary survey estimator is the simple mean
\begin{equation}
\hat{\mu}_{\mathrm{survey}}=\frac{1}{n}\sum_{i=1}^n Y_i.
\end{equation}
A prediction-only estimator, $N^{-1}\sum_{j=1}^N f(\tilde X_j)$, is not trustworthy by itself because the predictor can be biased. Prediction-powered inference keeps the prediction mean but adds a human-survey correction for prediction error:
\begin{equation}
\hat{\mu}_{\mathrm{PPI}}=
\frac{1}{N}\sum_{j=1}^N f(\tilde X_j)
+\frac{1}{n}\sum_{i=1}^n \{Y_i-f(X_i)\}.
\end{equation}
The first term estimates the predictable component of the outcome at auxiliary-file scale. The second term estimates the average residual that the predictor misses in the labeled survey. If the predictor is systematically too high, the residual correction is negative; if it is too low, the correction is positive. Taking expectations gives $E\{f(X)\}+E\{Y-f(X)\}=E(Y)=\mu$, so the estimator preserves the same target as the survey mean even when $E\{f(X)\}\neq E(Y)$. PPI can use biased predictions because the human survey estimates and corrects such bias.

The correction can often improve precision. Ignoring finite-population corrections and treating the survey and auxiliary samples as independent,
\begin{equation}
\mathrm{Var}(\hat{\mu}_{\mathrm{survey}})=\frac{1}{n}\mathrm{Var}(Y),
\qquad
\mathrm{Var}(\hat{\mu}_{\mathrm{PPI}})\approx
\frac{1}{N}\mathrm{Var}\{f(X)\}+\frac{1}{n}\mathrm{Var}\{Y-f(X)\}.
\end{equation}
When the auxiliary file is much larger than the survey, as ACS is here, the first term is negligible. PPI is then more efficient than the survey mean whenever the residual $Y-f(X)$ varies less than the original outcome $Y$. Put plainly, the survey no longer has to estimate the full outcome variation. It only has to estimate the part of the outcome that cannot be predicted.

The gain in precision can be captured intuitively by effective sample size. In the simple mean case, predictive accuracy links directly to that gain. If $f(X)$ is linearly calibrated so that the residual variance is approximately $\mathrm{Var}(Y)(1-R^2)$, then
\begin{equation}
\frac{\mathrm{se}(\hat{\mu}_{\mathrm{PPI}})}{\mathrm{se}(\hat{\mu}_{\mathrm{survey}})}
\approx \sqrt{1-R^2},
\qquad
n_{\mathrm{eff}}\approx \frac{n}{1-R^2}.
\end{equation}
A predictor with out-of-sample $R^2=0.10$ therefore gives the precision of a survey about $1/(1-0.10)=1.11$ times as large, or an 11 percent effective-sample-size gain. An $R^2$ of 0.20 gives a 25 percent gain, and an $R^2$ of 0.40 gives a 67 percent gain. These numbers explain why modest individual-level predictive accuracy can still be useful for survey inference when predictions are cheap, the auxiliary file is large, and collection of an actual data point is expensive.

The same unweighted construction extends from means to regression coefficients and other moment-defined estimands. For a target parameter $\theta^*$ satisfying
\begin{equation}
E\{\psi(Y,X;\theta^*)\}=0,
\end{equation}
define $\psi_f(X;\theta)=\psi\{f(X),X;\theta\}$. The unweighted PPI estimating equation is
\begin{equation}
U_{\mathrm{PPI}}(\theta)
=
\frac{1}{N}\sum_{j=1}^N \psi_f(\tilde X_j;\theta)
+
\frac{1}{n}\sum_{i=1}^n
\left[
\psi(Y_i,X_i;\theta)-\psi_f(X_i;\theta)
\right]
=0.
\end{equation}
This is the score-level analogue of the mean estimator above: prediction-implied scores are averaged over the large auxiliary file, and the labeled survey estimates the residual score correction. When $\psi(Y;\mu)=Y-\mu$, the equation reduces to the unweighted PPI mean. For linear regression, $\psi(Y,X;\beta)=X(Y-X^\top\beta)$ with an intercept included in $X$. For logistic or other generalized linear models, $\psi$ is the corresponding score.

Social surveys are usually not simple random samples. We therefore use a survey-weighted formulation of PPI. Weighted residual correction has been developed within inverse-probability-weighted PPI, and related prediction-assisted methods accommodate weighted, stratified, and clustered samples \citep{datta2025,kluger2025}. Let $S_\ell$ denote the labeled probability survey sample, with outcomes $Y_i$, covariates $X_i$, and design weights $w_i$. Let $S_u$ denote the auxiliary probability sample, with covariates $X_j$ and weights $a_j$ but without the focal outcome. Define normalized weights $\tilde{w}_i=w_i/\sum_{i\in S_\ell}w_i$ and $\tilde{a}_j=a_j/\sum_{j\in S_u}a_j$. The weighted mean estimator is
\begin{equation}
\hat{\mu}_{\mathrm{dcPPI}} = \sum_{j \in S_u} \tilde{a}_j f(X_j) + \sum_{i \in S_\ell} \tilde{w}_i\{Y_i - f(X_i)\}.
\label{eq:weighted-mean}
\end{equation}

The justification requires both files to represent the same target population, with compatible covariate measurements and weights appropriate to their respective sampling designs. Suppose that the weighted averages consistently estimate the relevant population averages. For a fixed external predictor, the auxiliary term then converges to $E\{f(X)\}$, while the survey residual term converges to $E\{Y-f(X)\}$. Their sum converges to $E(Y)$, regardless of whether the predictor is unbiased. Because the normalized weights produce ratio estimators, the claim is design consistency rather than exact finite-sample unbiasedness. The predictions can improve precision, but their bias correction remains grounded in observed survey outcomes.

The same correction applies to parameters defined by population moment conditions. The survey-weighted PPI estimator solves
\begin{equation}
U_{\mathrm{dcPPI}}(\theta) = \sum_{i\in S_\ell}\tilde{w}_i\psi(Y_i,X_i;\theta) + \left[\sum_{j\in S_u}\tilde{a}_j\psi_f(X_j;\theta) - \sum_{i\in S_\ell}\tilde{w}_i\psi_f(X_i;\theta)\right] = 0.
\label{eq:weighted-moment}
\end{equation}

Under the same population-alignment and weighting conditions, both weighted averages of $\psi_f(X;\theta)$ converge to the same population moment, so their difference converges to zero. Consequently, $U_{\mathrm{dcPPI}}(\theta)$ converges to $E\{\psi(Y,X;\theta)\}$. With uniform convergence of the weighted moments and an identified population solution, the corresponding estimator is consistent for $\theta^*$. This argument treats the predictor as fixed and external to the outcomes used for correction.

For coefficients and other score-based estimands, the relevant predictive signal is not individual-level $R^2$ alone. What matters is how much the predictor reduces the variation in the estimating score that the survey must estimate after rectification.

Uncertainty follows a first-order linearization of the estimating equation \citep{binder1983}. Let $A(\theta)=-\partial U_{\mathrm{dcPPI}}(\theta)/\partial\theta^\top$ and let $B(\theta)=\mathrm{Var}_d\{U_{\mathrm{dcPPI}}(\theta)\}$ denote its covariance under the sampling designs, holding $\theta$ and the external predictor fixed. The large-sample sandwich variance estimator is
\begin{equation}
\widehat{\mathrm{Var}}(\hat{\theta}_{\mathrm{dcPPI}})=\hat A^{-1}\hat B\hat A^{-\top},
\qquad
\hat A=A(\hat{\theta}_{\mathrm{dcPPI}}).
\label{eq:weighted-variance}
\end{equation}
Here $\hat B$ is a design-based covariance estimate constructed from score contributions evaluated at the fitted parameter. It is not the literal variance of the solved equation $U_{\mathrm{dcPPI}}(\hat{\theta}_{\mathrm{dcPPI}})=0$. The online supplement (Section~\ref{app:variance-proof}) states the regularity conditions and provides the detailed derivation of Equation \ref{eq:weighted-variance}.

To estimate $\hat B$, the survey contributes residual scores $\psi(Y_i,X_i;\theta)-\psi_f(X_i;\theta)$, and the auxiliary sample contributes prediction-implied scores $\psi_f(X_j;\theta)$. Taylor linearization or replicate weights must respect each sampling design, including weight normalization, strata, and clusters where applicable. The two covariance contributions add when the sampling designs are independent; otherwise, their cross-covariance must also be included. Treating auxiliary uncertainty as negligible is an approximation justified by its variance contribution relative to that of the survey residual correction, not by file size alone.

The main efficiency quantities are confidence-interval width reduction and effective-sample-size gain. Let $\widehat{se}_{\mathrm{survey}}$ be the standard error of the ordinary weighted survey estimator and $\widehat{se}_{\mathrm{dcPPI}}$ the standard error of the augmented estimator. The effective-sample multiplier is
\begin{equation}
\left(\frac{\widehat{se}_{\mathrm{survey}}}{\widehat{se}_{\mathrm{dcPPI}}}\right)^2,
\end{equation}
and the ESS gain is this multiplier minus one. A 10 percent ESS gain means that the augmented estimator has the precision of a survey-only estimator with roughly 10 percent more observations, holding the design otherwise fixed. We also report confidence-interval width reduction as $1-W_{\mathrm{dcPPI}}/W_{\mathrm{survey}}$, where $W$ is the 95 percent interval width. 

\subsection{Prediction engines}

The benchmark uses three prediction engines, each designed to produce predictions external to the outcome values being rectified. The first is historical supervised learning. For each target outcome, a Super Learner ensemble is trained on historical survey waves and then applied to the target-year survey and auxiliary file \citep{vanderlaan2007}. The learner library contains generalized linear models, elastic-net regularization, and random forests. Historical training is strongest when survey instruments are stable across time and when prior waves contain many labeled examples. It is fully out-of-time relative to the target wave, so the predictions used for rectification are not fitted on the target outcomes.

The second regime is within-wave cross-fitting. The target-year survey is partitioned into folds; models are trained on all but one fold and then used to predict the held-out fold. Each labeled unit's prediction is therefore generated by a model that did not use that unit's outcome. This reduces direct in-sample overfitting and follows the logic of cross-prediction-powered inference \citep{zrnic2024}. It is useful when historical training data are unavailable or when temporal drift makes historical prediction suspect. After out-of-fold predictions are obtained for the labeled survey, a final model is trained on the full target-year survey to score the auxiliary ACS file. Extending the fixed-predictor variance argument to this regime requires additional stability conditions and agreement between the fold-trained and full-wave predictors at the first-order scale; held-out predictions alone do not establish these conditions (see the online supplement, Section~\ref{app:variance-proof}).

\begingroup
\widowpenalty=10000
The third regime is LLM respondent simulation. Each respondent's covariates are rendered as a structured demographic profile, paired with the verbatim survey question and admissible response options, and passed to a language model with a constrained output format. Prompted models include GPT-5-mini, Gemini-2.5-flash, and Qwen3, and one Qwen3 variant is fine-tuned on historical respondent--outcome pairs. Prompting and fine-tuning address the same prediction task, which is a survey response conditional on a respondent's demographic profile and the question being asked. In-context learning supplies observed examples within the prompt, while supervised fine-tuning learns from observed respondent--outcome pairs. In both approaches, the final predicted response provides auxiliary predictive information for the inference analysis. More implementation details can be found in the online supplement (Section~\ref{app:llm-implementation}).\par
\endgroup

\section{Data}

The auxiliary frame is the American Community Survey as harmonized by IPUMS USA. IPUMS USA provides annual ACS public-use microdata for the period since 2000 and harmonizes demographic, household, and economic variables across years. We use ACS 2017 as the auxiliary frame for the NHIS benchmark and ACS 2022 for the GSS benchmark. In the processed files used for the analyses, ACS 2017 contributes roughly 3.19 million person records and ACS 2022 contributes roughly 3.37 million person records. Although we describe ACS as census-scale, the analysis uses ACS public-use microdata rather than a complete census. We treat ACS as a large weighted auxiliary file and align it to the target survey populations as closely as possible using age, household, and person-level restrictions. For the GSS benchmark, the target is U.S. adults in the 2022 GSS universe; for the NHIS benchmark, the target is the civilian noninstitutionalized population represented by NHIS 2017. ACS person weights are used for auxiliary prediction means, while GSS and NHIS design weights define the labeled-survey residual correction.

The opinion outcomes come from the 2022 General Social Survey, a national survey of U.S. adults conducted by NORC. The public-use GSS file contains 3,544 respondents in the 2022 target wave, and historical waves from 1988 through 2021 provide 46,971 observations for training and prompt exemplars \citep{davern2024}. We selected 37 opinion-related outcomes from GSS, with details in the online supplement (Section~\ref{app:gss-outcomes}). These outcomes roughly fall under three categories: political attitudes, social attitudes, and well-being and quality of life. 

The health outcomes come from the 2017 National Health Interview Survey, a major U.S. health survey conducted by the National Center for Health Statistics. NHIS is designed to monitor the health of the civilian noninstitutionalized population and to relate health characteristics to demographic and socioeconomic variables \citep{nchs2017}. The target wave contains 78,132 observations in the processed benchmark, and pre-2017 waves provide a large historical training corpus of more than six million observations. We selected 46 health-related outcomes, with details in the online supplement (Section~\ref{app:nhis-outcomes}). These outcomes roughly fall under three categories: health behavior, health status, and mental health. We use NHIS 2017 because more recent waves contain more missingness in some harmonized demographic covariates needed for ACS alignment. 

For both GSS and NHIS, binary outcomes are coded 0/1, ordered categorical outcomes retain their substantive ordering, and count or continuous outcomes are analyzed on their released numeric scale after invalid and not-in-universe codes are removed. The benchmark excludes unordered categorical outcomes with more than two categories. This restriction applies only to the outcomes being predicted and analyzed, not to the covariates; nominal categorical covariates, such as race, are included among the predictors. LLM predictions are parsed as valid response labels and then mapped to the same numeric codes used for survey responses.

\placehere{Table~\ref{tab:data-sources}}

The predictor space is deliberately limited to variables that can be represented in both the target surveys and ACS. The harmonized covariates include sex, age, marital status, race, nativity, citizenship, educational attainment, employment status, occupation, industry, personal income, and poverty status. Occupation and industry are recoded to common census-based categories; income and education are harmonized to comparable ordered categories; and special missing, refused, not-in-universe, and invalid codes are excluded before modeling or estimation. Survey design variables, including person or household weights, strata, and primary sampling units where available, are retained for design-consistent estimation.

The outcome set is intentionally heterogeneous. This heterogeneity is useful because prediction-powered gains should not be expected to be uniform. Health status, activity limitation, and home ownership are plausibly more tied to demographic and socioeconomic covariates than many political opinions or moral attitudes.

\section{Results}

The main results proceed in three steps. We first pair GSS and NHIS outcomes with ACS auxiliary covariates to compare expected gains for means and regression coefficients. We then examine when LLM predictors may be useful under label scarcity and how PPI changes hypothesis-testing behavior in small samples. The online supplement (Section~\ref{app:supplementary-results}) provides supplementary analyses of repeated-sampling performance and the differences between prediction-only, PPI, and survey-only point estimates.

\subsection{Predictability and expected efficiency gain using GSS and NHIS data}

The first analysis pairs NHIS and GSS outcomes with ACS auxiliary covariates to evaluate expected PPI efficiency gains for two types of inferential targets: survey-weighted outcome means and regression coefficients. For each outcome, we specify a survey-weighted linear regression with an intercept and sex, age, education, and race entered jointly. We examine four coefficients: female relative to male, age in years, the harmonized education scale, and White relative to the non-White reference group. These coefficients describe demographic associations adjusted for the other covariates and allow us to assess whether PPI improves precision for associations as well as means. The prediction engines described in Section 3.2 supply the outcome predictions used to augment estimation of these targets.

Table \ref{tab:summary-gains} averages the same outcome-estimand panels used in Figures \ref{fig:nhis-eff} and \ref{fig:gss-eff}. Each cell gives confidence-interval width reduction followed by effective-sample-size gain. For NHIS means, the historical learner reduces interval width by 5.4 percent and increases effective sample size by 12.2 percent, while the cross learner yields 4.3 percent and 9.8 percent. For GSS means, the corresponding gains are 3.4 percent and 7.5 percent for the historical learner and 3.0 percent and 6.5 percent for the cross learner. The best LLM rows are positive but smaller, with mean-level gains of 1.2 percent and 2.6 percent in NHIS and 1.4 percent and 2.8 percent in GSS. Coefficient targets are less uniform and usually smaller, especially in the GSS supervised rows, because prediction must reduce the variation in the estimating contributions relevant to each coefficient. These gains are modest for any single estimate but meaningful at survey scale, where a widely used public survey effectively becomes larger for many estimands without additional interviews.

\placehere{Table~\ref{tab:summary-gains}}

Table \ref{tab:summary-gains} summarizes average gains across outcomes. Figure \ref{fig:nhis-eff} disaggregates these gains by outcome and estimand for NHIS, and Figure \ref{fig:gss-eff} presents the corresponding results for GSS.

In NHIS, the strongest mean-level gains occur for health status, functional limitations, and physical-activity measures. The largest mean effective-sample-size gains are roughly 34 to 36 percent for self-rated health, limitation measures, and minutes of vigorous physical activity. Smoking frequency and some mental-health intensity measures are less predictable from the harmonized covariates and yield smaller gains. In GSS, the largest mean gains are about 21 to 23 percent for computer use, marijuana legalization, home ownership, and social class. Home ownership, political ideology, gun ownership, and financial satisfaction carry useful predictive signal, while many social and political attitudes remain weakly predicted by demographic and socioeconomic covariates alone.

Mean estimation benefits when predictions reduce the outcome variation left in the survey residual correction. For a regression coefficient, gains depend on reductions in the variation of the estimating contributions relevant to that coefficient. In this benchmark, coefficient-level gains are usually smaller and more heterogeneous than mean-level gains. This distinction matters when the scientific target is an adjusted group difference or association rather than a marginal mean.

\placehere{Figure~\ref{fig:nhis-eff}}

\placehere{Figure~\ref{fig:gss-eff}}

\begin{figure}[p]
\centering
\includegraphics[width=0.95\textwidth]{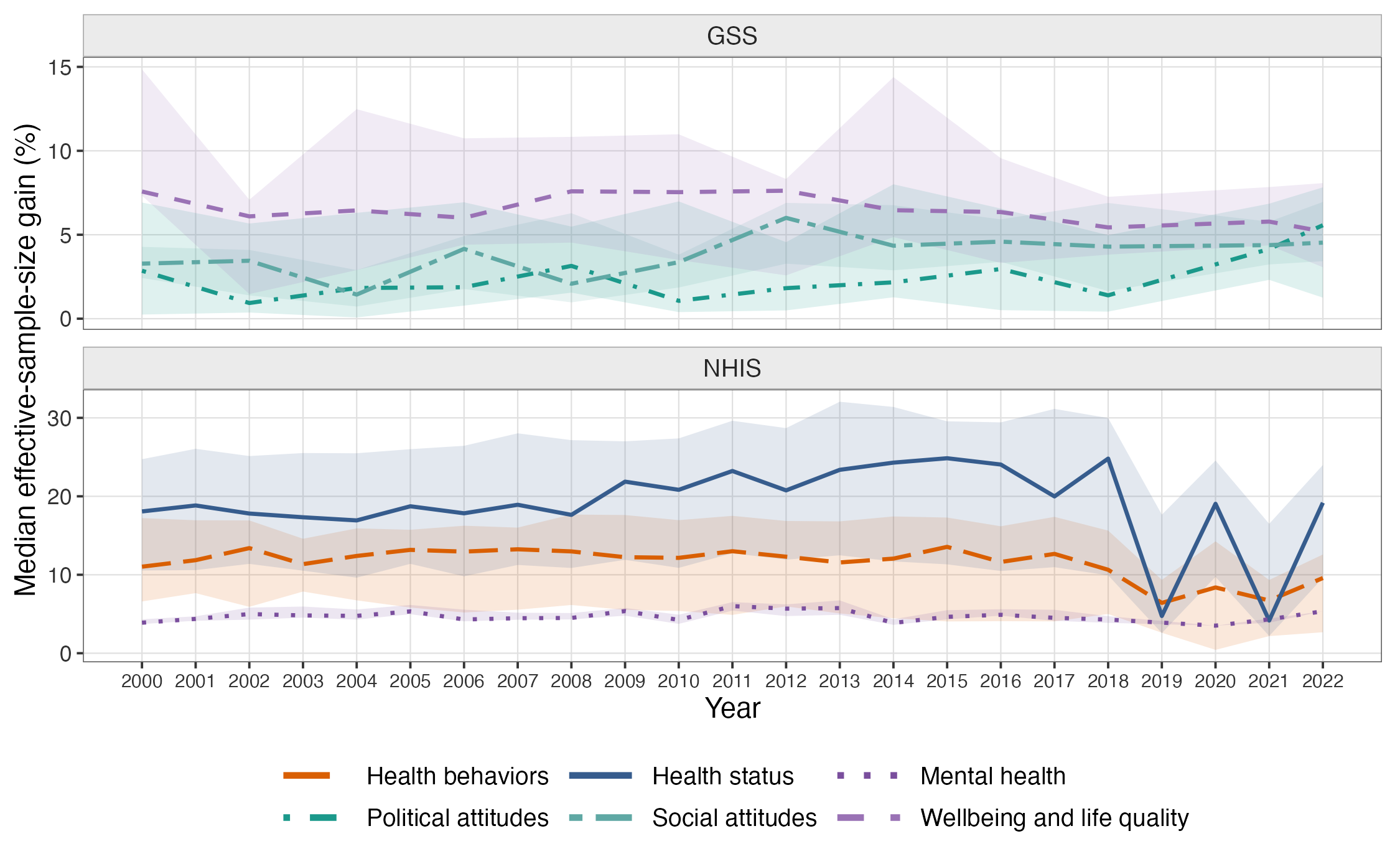}
\caption{Median effective-sample-size gains over time by outcome family. Health-related outcomes are generally more predictable from harmonized census-style covariates than political and social attitudes, though the magnitude varies across years and survey domains.}
\label{fig:temporal}
\end{figure}

Figure \ref{fig:temporal} extends this heterogeneity over time. Here we report results based on historical supervised learners. Across repeated survey years, NHIS health-status outcomes have median effective-sample-size gains ranging from about 4 to 25 percent, health behaviors range from about 6 to 14 percent, and mental-health outcomes range from about 4 to 6 percent. GSS political attitudes, social attitudes, and well-being measures mostly range from the low single digits to about 8 percent. This pattern does not imply that attitudes are intrinsically unpredictable. NHIS has a larger training corpus than GSS, which could explain part of the difference in predictability. Richer auxiliary data could also increase predictability. 

\placehere{Figure~\ref{fig:temporal}}

Table \ref{tab:summary-gains} and Figures \ref{fig:nhis-eff}--\ref{fig:temporal} summarize expected confidence-interval width reductions and effective-sample-size gains for the available ACS covariates. These comparisons show which outcomes and estimands are most likely to benefit from augmentation in this benchmark. Where expected gains are small, additional observed outcomes or richer auxiliary covariates may be more useful.

\subsection{When LLM predictions may still help}

The second analysis asks whether LLMs still have a role despite the stronger performance of supervised learners. The limitation of supervised learning is that it needs labeled outcome data, while LLM prediction does not. Historical and cross learners require labeled outcomes from prior waves or within-wave training folds, while an LLM can generate external predictions before a new survey item has accumulated training data.

\begin{figure}[p]
\centering
\includegraphics[width=0.75\textwidth]{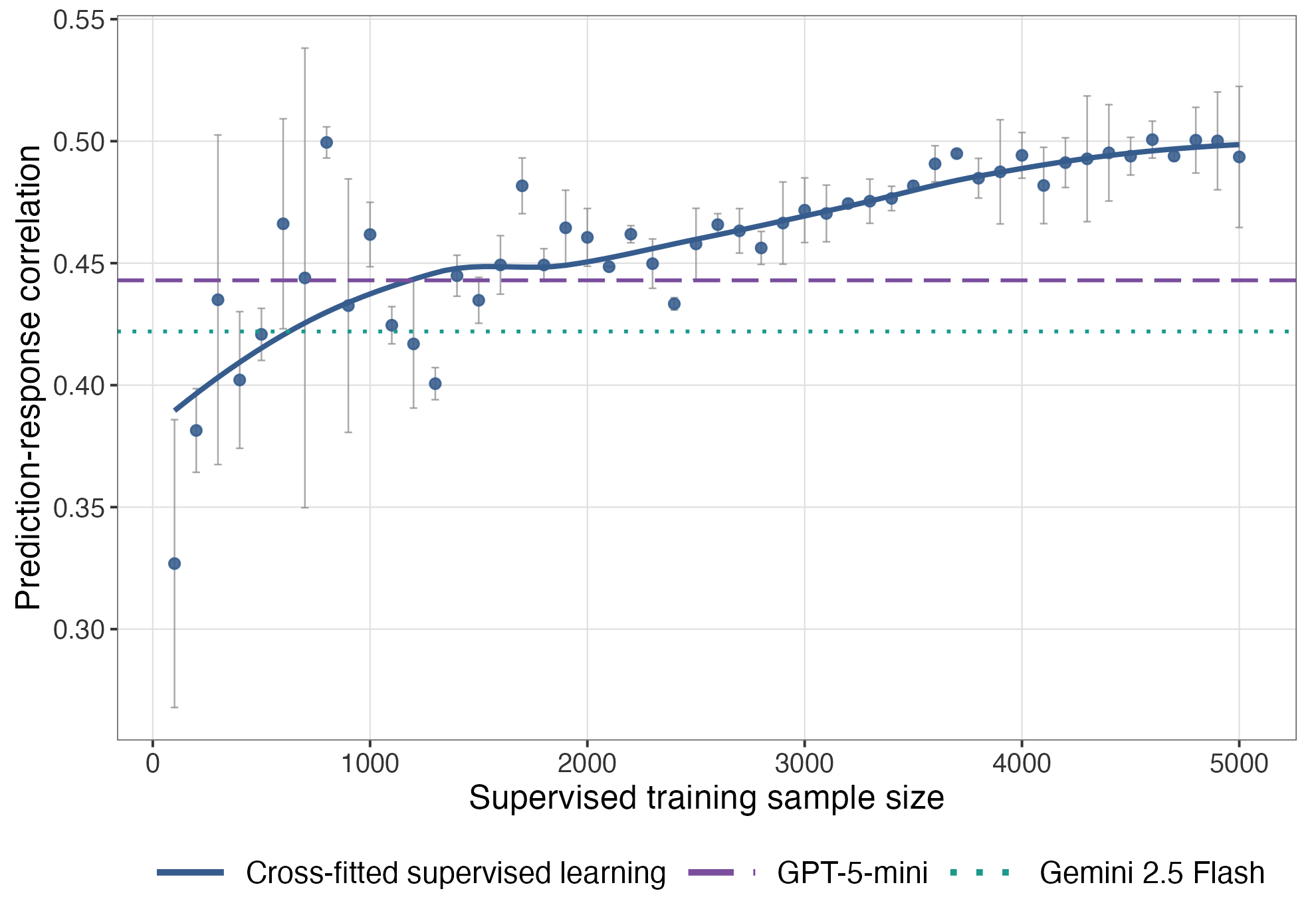}
\caption{Predictive accuracy for self-rated health. Cross-fitted supervised learning is compared with LLM respondent simulation. LLM predictions contain useful signal, especially relative to very small training samples, but supervised learning improves as labeled survey data accumulate.}
\label{fig:llm-ml}
\end{figure}

Figure \ref{fig:llm-ml} illustrates this role under limited training data for self-rated health. Cross-fitted supervised learning reaches a prediction-response correlation of about 0.33 with 100 training observations, about 0.46 with 1,000 observations, and about 0.49 with 5,000 observations. GPT-5-mini has a correlation of about 0.443, roughly comparable to the supervised learner near 1,000 training observations. Gemini 2.5 Flash is lower at about 0.422. As the supervised training sample grows, the data-based model continues to improve and eventually surpasses the LLM benchmarks. However, LLMs can be useful in the early stage of a measurement task, before outcome-specific training data are abundant.

\placehere{Figure~\ref{fig:llm-ml}}

\subsection{Small-sample implications for hypothesis testing}

\begin{figure}[p]
\centering
\begin{subfigure}{0.80\textwidth}
\centering
\includegraphics[width=\textwidth]{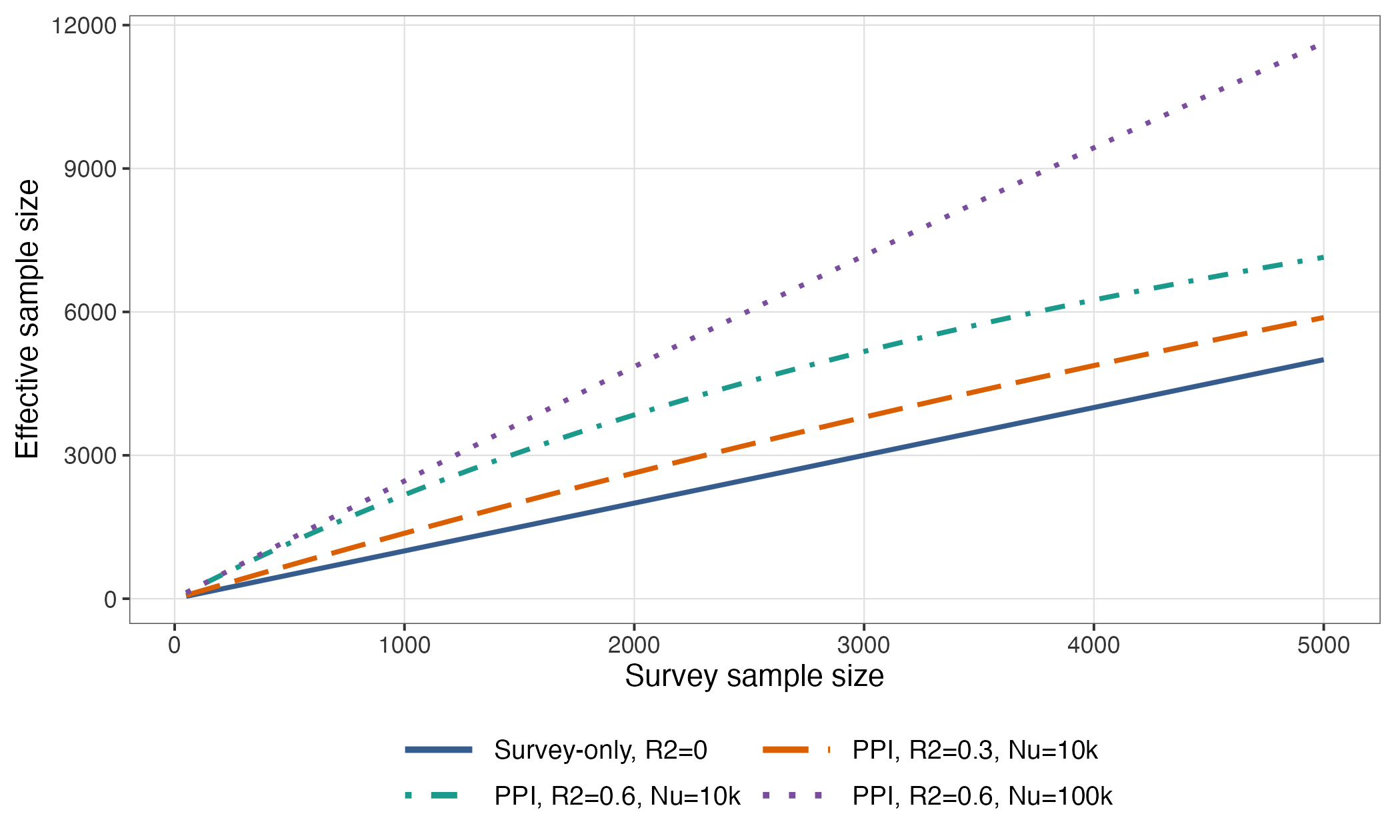}
\caption{Effective sample size}
\end{subfigure}
\vspace{0.5em}
\begin{subfigure}{0.80\textwidth}
\centering
\includegraphics[width=\textwidth]{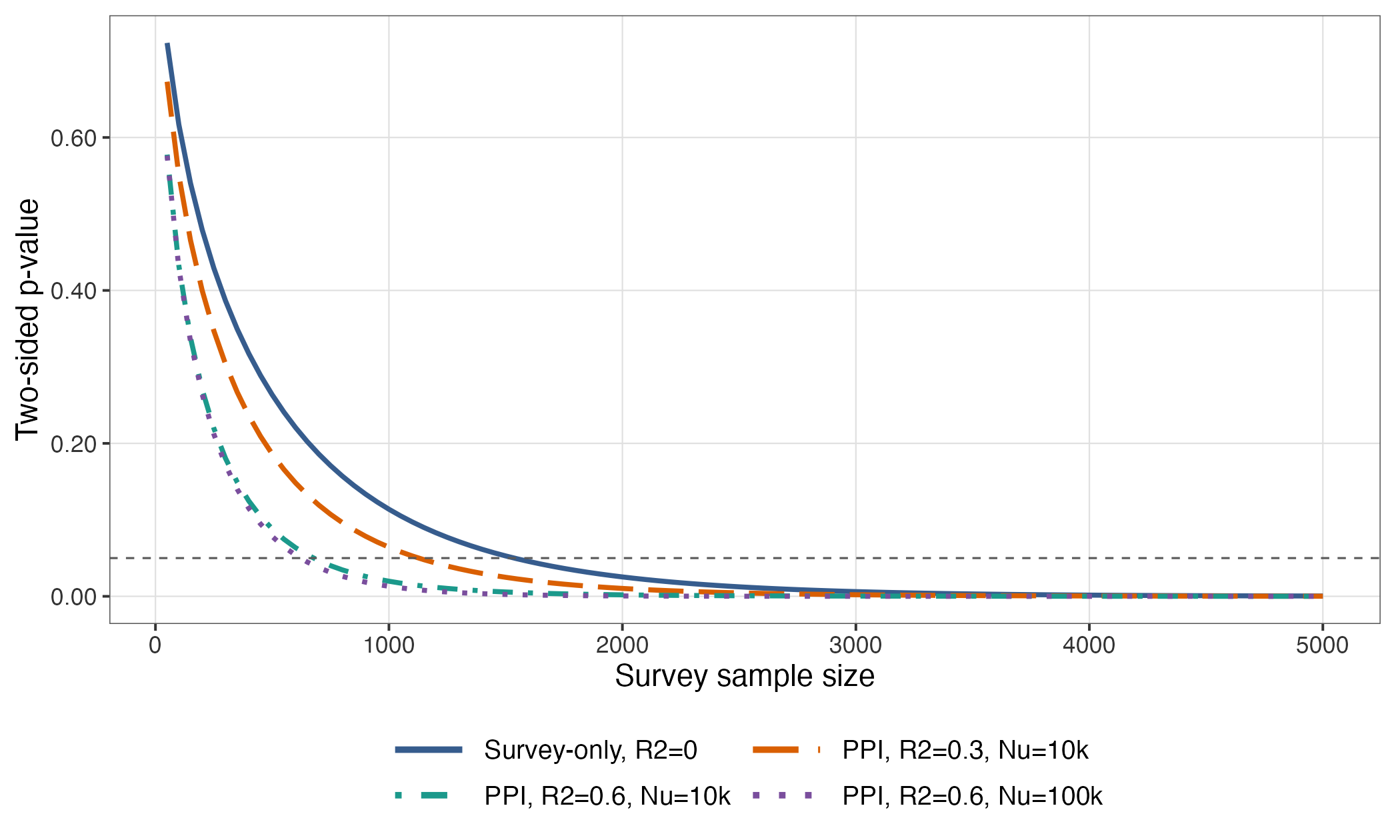}
\caption{Two-sided p-value}
\end{subfigure}
\caption{Small-sample implications of prediction-powered inference. In this analytic design illustration, PPI has greater consequences for the test conclusion when the survey-only calculation is near the significance threshold, the auxiliary file is large, and predictions substantially reduce residual outcome variation.}
\label{fig:small-sample}
\end{figure}

Small samples often arise within large surveys when researchers compare groups defined by intersecting social positions or local contexts. Imprecise group-specific estimates can obscure substantively important differences or encourage pooling categories that matter to the research question. Prediction-powered augmentation may improve precision when compatible auxiliary data and observed survey responses are available for the relevant groups. The final analysis illustrates the underlying mechanism with a mean tested against a fixed reference, holding the effect size fixed while sample size varies. Figure \ref{fig:small-sample} is an analytic design illustration, with the setup described in the online supplement (Section~\ref{app:small-sample-simulation}).

Figure \ref{fig:small-sample} describes the non-linearity in the gain for statistical tests. The upper panel translates predictive strength and auxiliary-file size into effective sample size. At a survey sample size of 1,000, survey-only analysis has an effective sample size of 1,000. PPI raises this to about 1,370 under a scenario with predictive $R^2=0.30$ and an auxiliary file with 10,000 observations. With predictive $R^2=0.60$, effective sample size rises to about 2,174 when the auxiliary file has 10,000 observations and about 2,463 when it has 100,000 observations. At a survey sample size of 5,000, the same scenarios yield effective sample sizes of about 5,882, 7,143, and 11,628. Given a predictive model and auxiliary data, effective sample size increases roughly linearly with actual sample size.

On the other hand, it is well known that the p-value does not change linearly with sample size. The lower panel gives the corresponding two-sided test behavior. At a survey sample size of 1,000, survey-only analysis gives \(p\approx0.114\). PPI with \(R^2=0.3\) and \(N_u=10{,}000\) lowers the p-value to about 0.064. PPI with \(R^2=0.6\) lowers it to about 0.020 when \(N_u=10{,}000\) and about 0.013 when \(N_u=100{,}000\). Conventional significance is reached at about 1,550 survey cases under survey-only analysis, about 1,150 cases with \(R^2=0.3\), about 700 cases with \(R^2=0.6\) and \(N_u=10{,}000\), and about 650 cases with \(R^2=0.6\) and \(N_u=100{,}000\). The largest changes in the illustrated test conclusion occur where the survey-only calculation lies near the significance threshold.

\placehere{Figure~\ref{fig:small-sample}}

For subgroup comparisons, predictive performance and the residual correction must be assessed within the relevant groups. Overall accuracy can conceal weak prediction for a small group. The variance of a group difference must incorporate uncertainty in both corrected group estimates and any covariance between them. Where group-specific predictions contain useful signal, augmentation may improve precision without pooling substantively distinct groups.

\section{Discussion}

The central claim of this article is deliberately conservative: AI predictions cannot replace social surveys, but they can augment them. AI predictions are too biased to be trusted, but are too informative to ignore. The empirical evidence supports both halves of the claim. Replacement fails because synthetic responses can be biased, and more synthetic observations can make that bias look deceptively precise. Augmentation works because predictions can contain useful information about outcomes, and design-consistent rectification can use that information without trusting it unconditionally.

The results identify where survey augmentation is most promising in this benchmark. Gains are generally larger for outcome means than regression coefficients and for health measures than social and political attitudes. Mean estimation benefits when predictions reduce outcome variation left for the survey to estimate; coefficient estimation requires predictive information relevant to the particular association being estimated. Researchers studying adjusted associations should therefore assess coefficient-specific gains rather than infer them from performance in predicting the outcome. Weakly predictable outcomes may offer limited returns from augmentation with the available covariates. Supervised learning provides the strongest gains when labeled training data are available, while LLMs may be useful for new survey items when such data are scarce. The practical question is how much precision the available predictions add for the researcher's intended estimand.

Our findings also encourage survey agencies and data producers to be prediction-aware. Prediction-aware design means that harmonized links to population-scale covariates and predictive algorithms can be released together with public-use files. Analysts could report both survey-only and augmented estimates, with precision gains evaluated separately for each estimand.

Prediction-aware design can also change fieldwork decisions. Historical predictability profiles can identify which outcomes are likely to benefit from auxiliary prediction and which require direct measurement. During data collection, early interviews can be used to assess whether a focal estimand has reached a target precision; if so, fieldwork can be shifted toward outcomes, modules, or groups where predictive signal is weak. In questionnaire design, agencies might keep a stable core of outcomes measured on all respondents while rotating or splitting additional modules, using PPI to recover precision for outcomes that are easier to predict and reserving scarce interview time for outcomes that are not. 

LLMs occupy a specific place in this future. They are useful as flexible external predictors, especially for new modules, rare outcomes, or early fieldwork when historical training data are unavailable. But their role should remain auxiliary. LLM outputs should be audited and compared against simple and supervised baselines. Most importantly, LLM predictions should be tied to a probability sample through rectification before they enter population inference.

PPI depends on the labeled survey and auxiliary file representing the same target population after weighting and population restrictions. Differences in coverage or covariate measurement can prevent the prediction terms from cancelling, and PPI does not itself correct unaccounted-for nonresponse or coverage bias in the survey. Precision gains are also conditional in the sense that weak or poorly scaled predictions can leave residual variance unchanged or increase it, and auxiliary-file uncertainty can offset part of the gain. Inaccurate historical predictions can still be corrected under the population-alignment conditions, but may contribute little precision. Very small labeled samples, subgroup cells, or numbers of independent sampling units can make residual corrections unstable and large-sample confidence intervals unreliable.

The benchmark also has limitations. It uses harmonized demographic and socioeconomic covariates available in ACS. Richer auxiliary data, such as geography, administrative records, neighborhood context, or prior panel responses, could increase gains, especially for opinions and behaviors weakly predicted by demographics alone. But richer data also create privacy, governance, and fairness concerns. The analysis focuses on the United States, where ACS provides unusually large public-use microdata; other countries will differ in the availability of census, registry, or administrative frames. The benchmark reports average gains across selected GSS and NHIS outcomes; other outcomes, surveys, and estimands may differ.

In our opinion, the best path forward for surveys is neither pretending that the data environment has not changed nor abandoning human measurement for synthetic substitutes. The path is integrating human responses, auxiliary frames, and predictions in ways that preserve the inferential virtues of probability sampling. AI can help make surveys cheaper, more powerful, and more adaptive. It should do so by increasing the return on each human interview, not by erasing the need for interviews altogether.

\clearpage
\begingroup
\raggedright
\bibliographystyle{supporting_files/socsciauthor}
\bibliography{supporting_files/references}
\endgroup

\clearpage
\begin{table}[p]
\centering
\caption{Data sources and roles in the benchmark}
\label{tab:data-sources}
\begin{threeparttable}
\small
\setstretch{1.1}
\begin{tabularx}{\textwidth}{@{}l l l >{\raggedright\arraybackslash}X r@{}}
\toprule
Data source & Role & Years & Content & Sample size \\
\midrule
ACS/IPUMS & Auxiliary frame & 2017 & Covariates & 3,190,040 \\
ACS/IPUMS & Auxiliary frame & 2022 & Covariates & 3,373,378 \\
GSS & Target survey & 2022 & 37 outcomes + Covariates & 3,544 \\
GSS & Training corpus & 1988--2021 & Outcomes + Covariates & 46,971 \\
NHIS & Target survey & 2017 & 46 outcomes + Covariates & 78,132 \\
NHIS & Training corpus & 1963--2016 & Outcomes + Covariates & 6,141,631 \\
\bottomrule
\end{tabularx}
\begin{tablenotes}[flushleft]
\footnotesize
\item Notes: Sample sizes refer to processed benchmark files before outcome-specific missingness restrictions. Outcome-specific sample sizes vary because not every question is asked of every respondent and because invalid or not-in-universe codes are treated as missing.
\end{tablenotes}
\end{threeparttable}
\end{table}

\clearpage
\begin{table}[p]
\centering
\caption{Average efficiency gains from design-consistent PPI in the outcome-level panels}
\label{tab:summary-gains}
\begin{threeparttable}
\footnotesize
\setstretch{1.1}
\begin{tabularx}{\textwidth}{@{}l >{\raggedright\arraybackslash}p{1.25in} *{5}{>{\centering\arraybackslash}X}@{}}
\toprule
Dataset & Predictor & Mean & Female & Age & Education & White \\
\midrule
NHIS health & Historical learner & 5.4 / 12.2 & 2.9 / 6.4 & 2.6 / 5.6 & 3.8 / 8.3 & 3.2 / 7.0 \\
NHIS health & Cross learner & 4.3 / 9.8 & 2.1 / 4.6 & 1.8 / 3.8 & 2.6 / 5.8 & 2.3 / 5.0 \\
NHIS health & Best LLM (GPT-5-mini) & 1.2 / 2.6 & 0.1 / 0.3 & 0.2 / 0.4 & 0.3 / 0.6 & 0.3 / 0.7 \\
GSS opinions & Historical learner & 3.4 / 7.5 & 0.6 / 1.3 & 1.0 / 2.1 & 1.3 / 2.7 & 1.2 / 2.4 \\
GSS opinions & Cross learner & 3.0 / 6.5 & 0.5 / 1.0 & 0.5 / 1.0 & 1.0 / 2.1 & 0.6 / 1.2 \\
GSS opinions & Best LLM (GPT-5 CoT) & 1.4 / 2.8 & 0.6 / 1.3 & 0.8 / 1.7 & 0.6 / 1.3 & 1.4 / 3.3 \\
\bottomrule
\end{tabularx}
\begin{tablenotes}[flushleft]
\footnotesize
\item Notes: Each cell reports two percentages, confidence-interval width reduction followed by effective-sample-size gain. Female, age, education, and white refer to coefficients from regressions of each outcome on sex, age, education, and race. The summaries use the same outcome-estimand panels as Figures~\ref{fig:nhis-eff} and \ref{fig:gss-eff}; the NHIS panel omits the three weekly physical-activity variables excluded from Figure~\ref{fig:nhis-eff}. The best LLM row selects the LLM predictor with the largest mean-level effective-sample-size gain within each dataset.
\end{tablenotes}
\end{threeparttable}
\end{table}


\clearpage
\clearpage
\begin{landscape}
\begin{figure}[p]
\centering
\includegraphics[width=\linewidth,height=0.82\textheight,keepaspectratio]{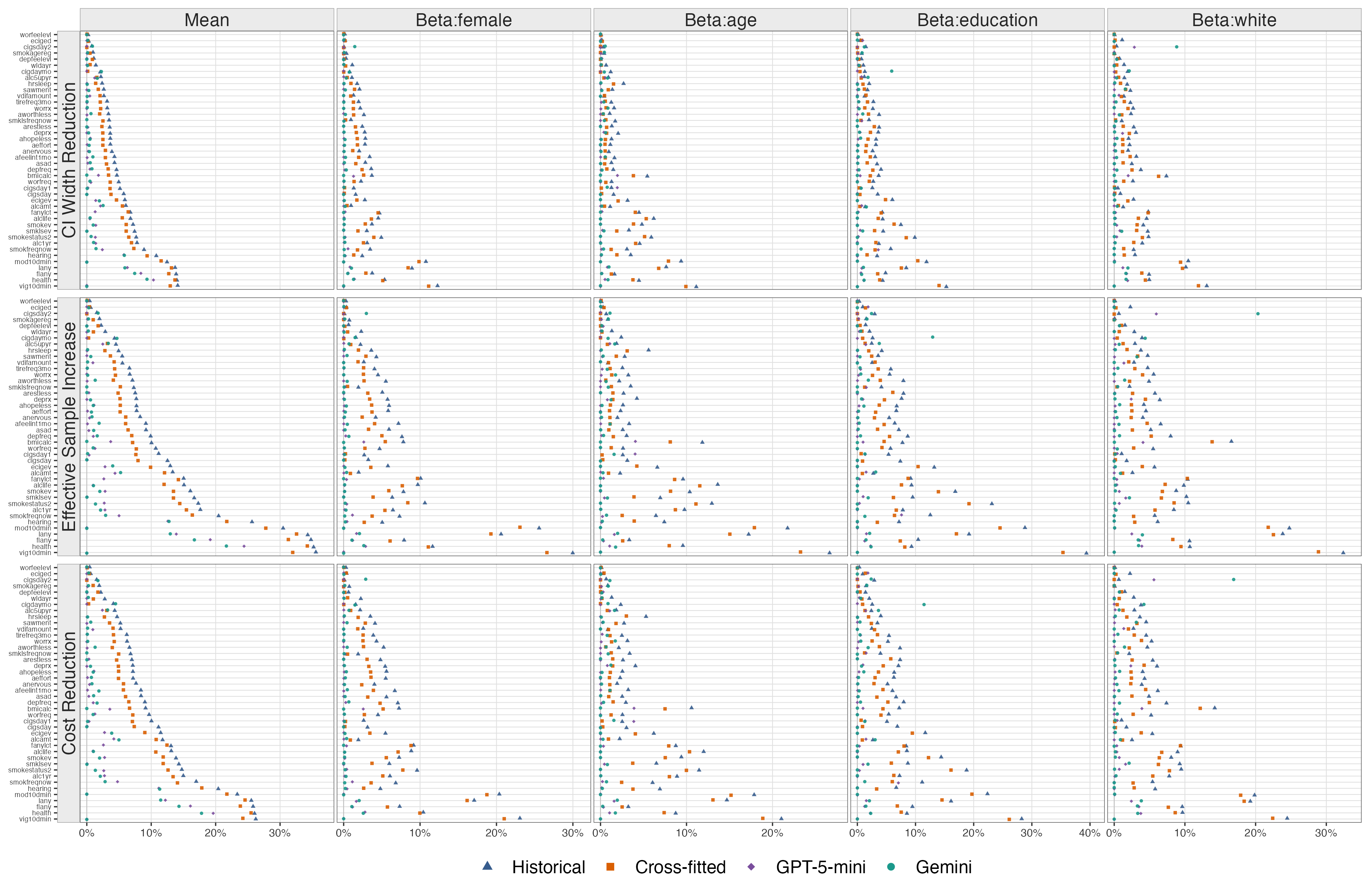}
\caption{Predictability and implied efficiency gains for NHIS 2017 health outcomes. The first column reports gains for survey-weighted means. The four coefficient columns report gains for terms from survey-weighted linear regressions of each outcome on sex, age, education, and race. These terms are female relative to male, age in years, the harmonized education scale, and White relative to the non-White reference category. Health status, functional limitations, hearing, and activity outcomes tend to yield larger gains than weakly predictable behavioral or mental-health intensity measures.}
\label{fig:nhis-eff}
\end{figure}
\end{landscape}
\clearpage
\begin{landscape}
\begin{figure}[p]
\centering
\includegraphics[width=\linewidth,height=0.80\textheight,keepaspectratio]{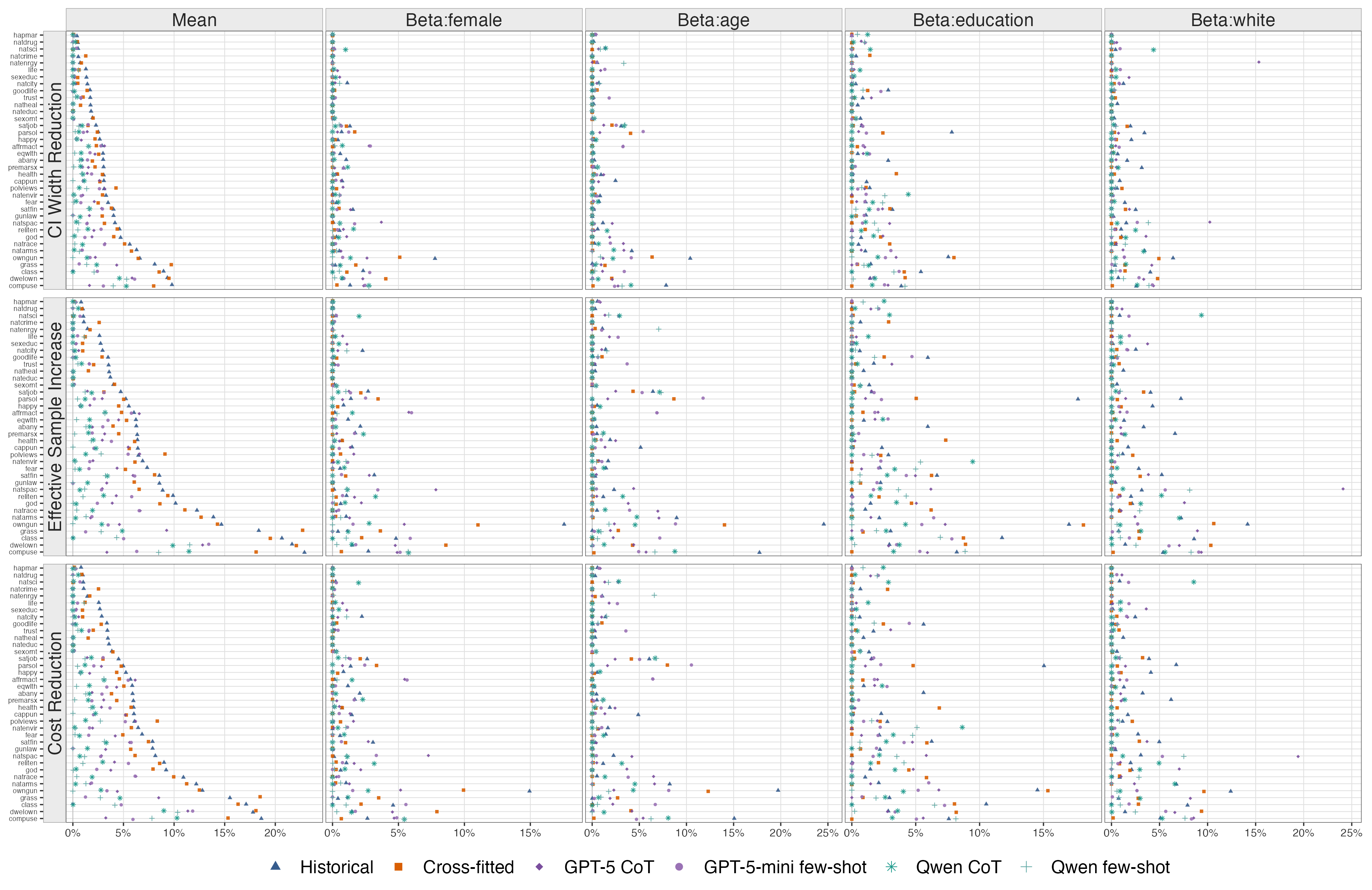}
\caption{Predictability and implied efficiency gains for GSS 2022 opinion and social-indicator outcomes. The first column reports gains for survey-weighted means. The four coefficient columns report gains for terms from survey-weighted linear regressions of each outcome on sex, age, education, and race. These terms are female relative to male, age in years, the harmonized education scale, and White relative to the non-White reference category. Qwen CoT coefficient results for \texttt{natenrgy} are omitted because four complete cases cannot identify five regression parameters. Gains are generally smaller than in NHIS and differ across outcomes and estimands, reflecting the weaker demographic predictability of many attitudes.}
\label{fig:gss-eff}
\end{figure}
\end{landscape}
\clearpage
\clearpage
\clearpage

\clearpage
\phantomsection
\addcontentsline{toc}{section}{Online Supplement}
\begin{center}
\vspace*{2em}
{\LARGE Online Supplement\par}
\vspace{1em}
{\large Using AI Predictions to Augment Rather Than Replace Surveys\par}
\vspace{1.5em}
Lai Wei \quad and \quad Yifei Zhu
\end{center}
\clearpage
\appendix
\onecolumn
\numberwithin{equation}{section}

\section{Statistical justification}
\label{app:statistical-justification}

\subsection{Derivation of the sandwich variance estimator}
\label{app:variance-proof}

This subsection derives Equation \ref{eq:weighted-variance} as a first-order asymptotic covariance estimator, rather than an exact finite-sample variance identity. The argument uses the estimating-equation linearization underlying complex-survey variance estimation \citep{binder1983}. Its scope is a fixed external predictor and sampling designs whose weighted moments and covariance estimates satisfy the conditions below. It does not require the prediction model to be correctly specified.

\paragraph{Target, notation, and conditions.}
For design-based inference, let $\mathcal P_m$ be a sequence of finite target populations and define
\begin{equation}
M_m(\theta)=\frac{1}{|\mathcal P_m|}\sum_{k\in\mathcal P_m}\psi(Y_k,X_k;\theta),
\qquad M_m(\theta_m^*)=0.
\label{eq:app-population-moment}
\end{equation}
The finite-population target $\theta_m^*$ is the counterpart of the expectation-defined target $\theta^*$ in Section 3.1 of the main article. Here $m$ indexes the sequence of populations and sampling designs, not an additional data source. Write $U_m(\theta)$ for Equation \ref{eq:weighted-moment} along this sequence, $A_m(\theta)=-\partial U_m(\theta)/\partial\theta^\top$, and $B_m(\theta)=\mathrm{Var}_d\{U_m(\theta)\}$. The subscript $d$ means that probability and covariance are taken over the joint sampling designs, conditional on the finite population and external predictor. The parameter dimension $p$ is fixed. The proof requires the following conditions:
\begin{enumerate}[label=(\roman*)]
\item Both samples refer to the same target population and use compatible covariates. Their positive design weights yield uniformly consistent weighted moment estimates on a neighborhood of the target. Score contributions have finite second moments, and the predictor is held fixed independently of the samples used for correction.
\item An identified interior root $\hat\theta_m$ of $U_m(\theta)=0$ is selected with $\hat\theta_m-\theta_m^*\xrightarrow{p_d}0$. The sample equation is continuously differentiable locally. Its negative derivative converges uniformly to the population negative derivative, which is uniformly continuous locally along the sequence and converges at $\theta_m^*$ to a nonsingular matrix $A_0$.
\item For a deterministic scale $r_m\to\infty$, a centered design central limit theorem holds:
\[
r_m U_m(\theta_m^*)\ \xrightarrow{d_d}\ \mathcal N_p(0,\Omega),
\qquad
r_m^2 B_m(\theta_m^*)\longrightarrow\Omega,
\]
where $\Omega$ is a finite covariance matrix. Any ratio-estimator bias is negligible at this scale. The scale reflects the sampling designs and relative sample sizes; it need not be the square root of the raw survey sample size.
\item A design-based score covariance estimator $\hat B_m$, evaluated with the fitted parameter held fixed, satisfies $r_m^2\hat B_m\xrightarrow{p_d}\Omega$. This condition requires a variance procedure valid for the actual sampling designs and normalized weights; positive weights alone do not establish it.
\end{enumerate}
Section 3.1 of the main article explains why the weighted moments have the correct population limit; condition (ii) makes consistency of the selected root explicit. Conditions (ii)--(iv) are the additional assumptions needed for the variance result. The derivation proceeds by linearizing the root, applying the design central limit theorem, and replacing the derivative and score covariance by consistent estimates.

\paragraph{Linearization of the root.}
Define the integrated negative derivative along the segment joining the target and fitted parameter,
\begin{equation}
\bar A_m=-\int_0^1
\frac{\partial U_m\{\theta_m^*+t(\hat\theta_m-\theta_m^*)\}}
{\partial\theta^\top}\,dt.
\label{eq:app-integrated-derivative}
\end{equation}
By consistency, that segment lies in the differentiable neighborhood with design probability tending to one. The fundamental theorem of calculus then gives the exact local identity
\[
0=U_m(\hat\theta_m)
=U_m(\theta_m^*)-\bar A_m(\hat\theta_m-\theta_m^*).
\]
Uniform derivative convergence, local uniform continuity, and consistency imply $\bar A_m\xrightarrow{p_d}A_0$. Because $A_0$ is nonsingular, $\bar A_m$ is invertible with probability tending to one. Hence
\begin{equation}
\hat\theta_m-\theta_m^*
=\bar A_m^{-1}U_m(\theta_m^*)
=A_0^{-1}U_m(\theta_m^*)+o_{p_d}(r_m^{-1}).
\label{eq:app-root-expansion}
\end{equation}
The remainder follows because $\bar A_m^{-1}-A_0^{-1}=o_{p_d}(1)$ and the central limit theorem implies $U_m(\theta_m^*)=O_{p_d}(r_m^{-1})$. Slutsky's theorem therefore yields
\begin{equation}
r_m(\hat\theta_m-\theta_m^*)
\ \xrightarrow{d_d}\
\mathcal N_p\!\left(0,A_0^{-1}\Omega A_0^{-\top}\right).
\label{eq:app-root-clt}
\end{equation}
This identifies the first-order asymptotic covariance. With $\hat A_m=A_m(\hat\theta_m)$, the same derivative conditions give $\hat A_m\xrightarrow{p_d}A_0$. Condition (iv) and continuity of matrix inversion imply
\[
r_m^2\hat A_m^{-1}\hat B_m\hat A_m^{-\top}
\ \xrightarrow{p_d}\
A_0^{-1}\Omega A_0^{-\top}.
\]
Removing the sequence subscripts gives Equation \ref{eq:weighted-variance}. This proves consistency for the first-order covariance under the stated conditions, not exact finite-sample unbiasedness of either the estimator or its estimated variance.

\paragraph{Design covariance of the score contributions.}
At a fixed parameter, decompose the sample equation as $U_m(\theta)=F_{u,m}(\theta)+R_{\ell,m}(\theta)$, where
\[
F_{u,m}(\theta)=\sum_{j\in S_u}\tilde a_j\psi_f(X_j;\theta),
\qquad
R_{\ell,m}(\theta)=\sum_{i\in S_\ell}\tilde w_i
\{\psi(Y_i,X_i;\theta)-\psi_f(X_i;\theta)\}.
\]
Their design covariance satisfies
\begin{equation}
\begin{split}
B_m(\theta)={}&\mathrm{Var}_d\{F_{u,m}(\theta)\}
+\mathrm{Var}_d\{R_{\ell,m}(\theta)\}\\
&+\mathrm{Cov}_d\{F_{u,m}(\theta),R_{\ell,m}(\theta)\}
+\mathrm{Cov}_d\{R_{\ell,m}(\theta),F_{u,m}(\theta)\}.
\end{split}
\label{eq:app-score-covariance}
\end{equation}
Independent sampling designs make the two cross-covariance terms zero. Dependent designs, including designs in which one sample is nested within the other, require their joint covariance to be accounted for. A genuinely complete auxiliary census has no auxiliary sampling variance conditional on the population; ACS public-use microdata, by contrast, are a sample. Neglecting their contribution is justified only when it is small relative to the residual-score contribution.

Both $F_{u,m}$ and $R_{\ell,m}$ use normalized weights and are ratio estimators. Taylor linearization must therefore account for the estimated weight totals in their denominators, as well as strata and clusters where applicable. Alternatively, a design-valid replicate-weight procedure can recompute the two normalized weighted averages in each replicate, holding $\hat\theta_m$ and the external predictor fixed. Their replicate sums estimate the covariance of the score contributions at that fixed parameter. Computing the variance of $U_m(\hat\theta_m)$ after resolving the root within every replicate would instead return zero and is not the construction of $\hat B_m$.

As a check, for mean estimation $\psi(Y;\mu)=Y-\mu$ and $\psi_f(X;\mu)=f(X)-\mu$, so $R_{\ell,m}$ is the weighted mean of $Y-f(X)$ and $\hat A_m=1$. Equation \ref{eq:weighted-variance} then reduces to the design variance of the prediction mean plus the residual correction, including their covariance when necessary. With independent simple random samples, negligible sampling fractions, and equal weights, this reproduces the unweighted variance expression in Section 3.1 of the main article.

\paragraph{Scope for learned predictors.}
The argument above conditions on an external predictor. If predictions are learned from the labeled survey, an additional first-order equivalence argument is required. A sufficient condition is
\[
r_m\{U_m^{\mathrm{learned}}(\theta_m^*)-
U_m^{\mathrm{fixed}}(\theta_m^*)\}=o_{p_d}(1),
\]
for a suitable fixed-predictor reference equation, together with the corresponding derivative and covariance consistency conditions. Cross-prediction uses fold-trained predictors on both the held-out labeled fold and the auxiliary sample and imposes learner-stability conditions \citep{zrnic2024}. When a full-wave model instead scores the auxiliary file, equivalence between that model and the fold-trained predictions must also be justified at the first-order scale. Merely excluding a respondent's own outcome from its prediction does not by itself prove Equation \ref{eq:weighted-variance} for this learned-predictor regime.

\clearpage
\section{LLM implementation details}
\label{app:llm-implementation}
\label{app:Implementation details}
This section describes the LLM survey-response prediction procedures, including in-context learning and supervised fine-tuning. Each prediction case combines a respondent's harmonized demographic profile, a survey question and its admissible responses. The final predicted response is mapped to the survey's numeric coding for the statistical analysis. Supervised fine-tuning uses GSS and NHIS respondent--outcome examples under this common prediction task.

\subsection{Experimental settings}
\label{app:llm-settings}
In-context examples pair respondent demographic profiles with observed survey responses. The GSS few-shot specification uses 20 historical examples per outcome from 2000--2021. Examples are sampled without replacement from respondents with an observed answer to the focal question, and the same outcome-specific set is used across target respondents. The NHIS few-shot specification uses 15 examples. Prompted chain-of-thought configurations allow intermediate reasoning, while the fine-tuning targets contain only the observed survey response.

Temperature controls variation in generated responses. We use temperature \(0.6\) for both in-context learning and inference from the fine-tuned model. This setting permits response variation, including intermediate reasoning in the chain-of-thought configurations. GSS negative special codes (e.g., \texttt{-100} to \texttt{-97}) and NHIS NIU/high missing codes (e.g., \texttt{0} or \texttt{996--999}) are treated as missing in unified files. Survey design weights are normalized in the estimating equation.

\subsection{Prompt templates and consistency across settings}
\label{app:llm-prompts}
\paragraph{Shared prediction task.}
Across in-context learning and supervised fine-tuning, the input combines a standardized demographic profile with a survey question and its valid response options. Each respondent--outcome pair contributes one final predicted response to the analysis. The response is mapped to the numeric coding used for the observed survey outcome. Chain-of-thought prompting allows intermediate reasoning, whereas the supervised fine-tuning targets contain response labels without reasoning annotations. Only the final predicted response enters the statistical analysis.

\paragraph{ICL (few-shot with reasoning): System template}
\begin{quote}\small
\begin{verbatim}
You are simulating a survey respondent. Answer exactly as instructed.

1. Analyze the demographic profile and the question
   to determine the most likely response.
2. Then output the final answer on the last line.

### Output format (strict)
PREDICTION: <verbatim answer>
(Use exactly one valid option label/number.
 No extra text after the PREDICTION line.)

As you answer, consider how similar respondents answered this question:

[FEW-SHOT EXEMPLARS: K examples for this outcome]
\end{verbatim}
\end{quote}

\begin{quote}\small
\begin{verbatim}
# Assistant target (SFT label)
PREDICTION: <verbatim answer>
\end{verbatim}
\end{quote}

\subsection{SFT training recipe (Qwen3-14B, LoRA)}
\label{app:llm-sft}
\paragraph{Model and objective.}
We fine-tune \texttt{Qwen3-14B} to map \((\text{profile},\text{question})\) to a \emph{single-line} prediction \texttt{PREDICTION: ...}. Training minimizes next-token cross-entropy on the assistant target under a chat-style \texttt{messages} format with System/User/Assistant roles.

\paragraph{Training data construction.}
Each training instance is one respondent--outcome pair.
The \textbf{System} input provides the survey-aware persona framing, enforces a strict no-reasoning instruction, and specifies the output-format constraint. The \textbf{User} input contains a standardized demographic profile and exactly one outcome definition with enumerated valid options. The \textbf{Assistant} target is exactly one line of the form \texttt{PREDICTION: <verbatim answer>} with no additional tokens.
To keep the distribution balanced without collapsing outcome coverage, we downsample \emph{respondents} rather than outcomes and mix GSS and NHIS examples under the unified predictor schema.

\paragraph{Train/validation split.}
We use \emph{year-disjoint} splits to mitigate leakage and to make validation reflect a genuine temporal holdout rather than random re-partitioning within the same wave. Concretely, we build the SFT dataset from postprocessed, harmonized survey files and define non-overlapping year windows separately for GSS and NHIS:
for \textbf{GSS}, we use train years 2000--2021; for \textbf{NHIS}, we use train years 2000--2016.
The benchmark-year inference tasks reported in the main paper are evaluated on held-out survey waves (e.g., the 2022 benchmark when used) and are not included in the SFT training window.

\paragraph{Sample selection.}
We select respondents before expanding their observed answers into respondent--outcome training examples. This controls the representation of each survey while retaining the questions answered by the selected respondents. Sample selection determines the composition of the training material; the training budget determines how much of that material is processed during fine-tuning.

\begingroup
\widowpenalty=10000
\paragraph{Mixing GSS and NHIS.}
Both surveys use a harmonized demographic representation and a common question--response prediction task. Pooling was intended to draw on the generalization capacity of LLMs and allow useful response patterns to transfer across survey domains. We control source representation at the respondent level so that the size of the NHIS respondent pool does not dominate the training mixture. We did not compare pooled and survey-specific models, so their relative predictive performance in this application remains unknown.\par
\endgroup

\begin{samepage}
\paragraph{LoRA configuration.}
We use parameter-efficient fine-tuning with LoRA, with a typical starting point of rank 32 and learning rate \(2\times 10^{-5}\), training for 1 epoch (formatting-focused tasks can overfit quickly). We set max sequence length to 2048 (increasing it if prompts truncate) and use AdamW as the optimizer.

\end{samepage}

\paragraph{Response format and inference.}
The supervised fine-tuning targets consist of observed survey response labels without reasoning annotations. Answer-only instructions request responses in this format. Prompted chain-of-thought configurations allow intermediate reasoning. For both approaches, only the final predicted response is used in the statistical analysis. Fine-tuned-model inference uses temperature \(0.6\), as described in Section~\ref{app:llm-settings}.

\clearpage
\section{Variable lists}
\label{app:variables}

\subsection{Harmonized predictors}
\small
\begin{longtable}{@{}>{\raggedright\arraybackslash}p{0.23\textwidth}>{\raggedright\arraybackslash}p{0.20\textwidth}>{\raggedright\arraybackslash}p{0.51\textwidth}@{}}
\toprule
\textbf{Variable} & \textbf{Category} & \textbf{Original question} \\
\midrule
\endfirsthead
\toprule
\textbf{Variable} & \textbf{Category} & \textbf{Original question} \\
\midrule
\endhead
\texttt{\detokenize{survey_year}} & Survey metadata & \detokenize{Survey year} \\
\texttt{\detokenize{sex}} & Demographics & \detokenize{Respondent sex} \\
\texttt{\detokenize{age}} & Demographics & \detokenize{Age in years} \\
\texttt{\detokenize{marital_status}} & Household / family & \detokenize{Marital status} \\
\texttt{\detokenize{race}} & Race / ethnicity & \detokenize{Race} \\
\texttt{\detokenize{born_in_us}} & Nativity / citizenship & \detokenize{Born in United States} \\
\texttt{\detokenize{citizenship}} & Nativity / citizenship & \detokenize{U.S. citizenship status} \\
\texttt{\detokenize{education}} & Education & \detokenize{Educational attainment} \\
\texttt{\detokenize{employment_status}} & Labor market & \detokenize{Employment status} \\
\texttt{\detokenize{occupation}} & Occupation / industry & \detokenize{Occupation code} \\
\texttt{\detokenize{sei}} & Occupational status &  \\
\texttt{\detokenize{prestige}} & Occupational status &  \\
\texttt{\detokenize{industry}} & Occupation / industry & \detokenize{Industry code} \\
\texttt{\detokenize{personal_income}} & Income / poverty & \detokenize{Personal income} \\
\texttt{\detokenize{poverty_status}} & Income / poverty & \detokenize{Poverty status} \\
\texttt{\detokenize{pid}} & Identifiers &  \\
\texttt{\detokenize{year_pid}} & Identifiers &  \\
\texttt{\detokenize{person_number}} & Identifiers & \detokenize{Person number within household} \\
\texttt{\detokenize{respondent_id}} & Identifiers & \detokenize{Respondent identifier} \\
\texttt{\detokenize{survey_weight}} & Survey design (weights) &  \\
\texttt{\detokenize{perweight}} & Survey design (weights) &  \\
\texttt{\detokenize{sampweight}} & Survey design (weights) &  \\
\texttt{\detokenize{fweight}} & Survey design (weights) &  \\
\texttt{\detokenize{hhweight}} & Survey design (weights) &  \\
\texttt{\detokenize{longweight}} & Survey design (weights) &  \\
\texttt{\detokenize{partweight}} & Survey design (weights) &  \\
\texttt{\detokenize{strata}} & Survey design (strata/PSU) & \detokenize{Sampling strata} \\
\texttt{\detokenize{psu}} & Survey design (strata/PSU) & \detokenize{Primary sampling unit} \\
\bottomrule
\end{longtable}
\normalsize

\subsection{GSS outcomes}
\label{app:gss-outcomes}
\small
\begin{longtable}{@{}>{\raggedright\arraybackslash}p{0.20\textwidth}>{\raggedright\arraybackslash}p{0.20\textwidth}>{\raggedright\arraybackslash}p{0.54\textwidth}@{}}
\toprule
\textbf{Variable} & \textbf{Category} & \textbf{Original question} \\
\midrule
\endfirsthead
\toprule
\textbf{Variable} & \textbf{Category} & \textbf{Original question} \\
\midrule
\endhead
\texttt{polviews} & Political Attitudes & \detokenize{We hear a lot of talk these days about liberals and conservatives. I'm going to show you a seven-point scale on which the political views that people might hold are arranged from extremely liberal---point 1---to extremely conservative---point 7. Where would you place yourself on this scale?} \\
\texttt{natspac} & Federal Government Spending & \detokenize{We are faced with many problems in this country, none of which can be solved easily or inexpensively. I'm going to name some of these problems, and for each one I'd like you to tell me whether you think we're spending too much money on it, too little money, or about the right amount. Space exploration program.} \\
\texttt{natenvir} & Federal Government Spending & \detokenize{We are faced with many problems in this country, none of which can be solved easily or inexpensively. I'm going to name some of these problems, and for each one I'd like you to tell me whether you think we're spending too much money on it, too little money, or about the right amount. Improving and protecting the environment.} \\
\texttt{natheal} & Federal Government Spending & \detokenize{We are faced with many problems in this country, none of which can be solved easily or inexpensively. I'm going to name some of these problems, and for each one I'd like you to tell me whether you think we're spending too much money on it, too little money, or about the right amount. Improving and protecting the nation's health.} \\
\texttt{natcity} & Federal Government Spending & \detokenize{We are faced with many problems in this country, none of which can be solved easily or inexpensively. I'm going to name some of these problems, and for each one I'd like you to tell me whether you think we're spending too much money on it, too little money, or about the right amount. Solving the problems of the big cities.} \\
\texttt{natcrime} & Federal Government Spending & \detokenize{We are faced with many problems in this country, none of which can be solved easily or inexpensively. I'm going to name some of these problems, and for each one I'd like you to tell me whether you think we're spending too much money on it, too little money, or about the right amount. Halting the rising crime rate.} \\
\texttt{natdrug} & Federal Government Spending & \detokenize{We are faced with many problems in this country, none of which can be solved easily or inexpensively. I'm going to name some of these problems, and for each one I'd like you to tell me whether you think we're spending too much money on it, too little money, or about the right amount. Dealing with drug addiction.} \\
\texttt{nateduc} & Federal Government Spending & \detokenize{We are faced with many problems in this country, none of which can be solved easily or inexpensively. I'm going to name some of these problems, and for each one I'd like you to tell me whether you think we're spending too much money on it, too little money, or about the right amount. Improving the nation's education system.} \\
\texttt{natrace} & Federal Government Spending & \detokenize{We are faced with many problems in this country, none of which can be solved easily or inexpensively. I'm going to name some of these problems, and for each one I'd like you to tell me whether you think we're spending too much money on it, too little money, or about the right amount. Improving the conditions of Black people.} \\
\texttt{natarms} & Federal Government Spending & \detokenize{We are faced with many problems in this country, none of which can be solved easily or inexpensively. I'm going to name some of these problems, and for each one I'd like you to tell me whether you think we're spending too much money on it, too little money, or about the right amount. The military, armaments and defense.} \\
\texttt{natsci} & Federal Government Spending & \detokenize{We are faced with many problems in this country, none of which can be solved easily or inexpensively. I'm going to name some of these problems, and for each one I'd like you to tell me whether you think we're spending too much money on it, too little money, or about the right amount. Supporting scientific research.} \\
\texttt{natenrgy} & Federal Government Spending &  \\
\texttt{eqwlth} & Inequality and Income & \detokenize{Some people think that the government in Washington ought to reduce the income differences between the rich and the poor, perhaps by raising the taxes of wealthy families or by giving income assistance to the poor. Others think that the government should not concern itself with reducing this income difference between the rich and the poor. Here is a card with a scale from 1 to 7. Think of a score of 1 as meaning that the government ought to reduce the income differences between rich and poor, and a score of 7 meaning that the government should not concern itself with reducing income differences. What score between 1 and 7 comes closest to the way you feel?} \\
\texttt{cappun} & Capital Punishment & \detokenize{Do you favor or oppose the death penalty for persons convicted of murder?} \\
\texttt{gunlaw} & Guns & \detokenize{Would you favor or oppose a law which would require a person to obtain a police permit before he or she could buy a gun?} \\
\texttt{grass} & Marijuana & \detokenize{Do you think the use of marijuana should be made legal or not?} \\
\texttt{reliten} & Religion & \detokenize{Would you call yourself a strong (PREFERENCE NAMED IN RELIG) or a not very strong (PREFERENCE NAMED IN RELIG)?} \\
\texttt{affrmact} & Affirmative Action & \detokenize{Some people say that because of past discrimination, Black people should be given preference in hiring and promotion. Others say that such preference in hiring and promotion of Black people is wrong because it discriminates against whites. What about your opinion -- are you for or against preferential hiring and promotion of Black people?} \\
\texttt{happy} & Happiness & \detokenize{Taken all together, how would you say things are these days--would you say that you are very happy, pretty happy, or not too happy?} \\
\texttt{hapmar} & Happiness / Marital Status & \detokenize{Taking things all together, how would you describe your marriage? Would you say that your marriage is very happy, pretty happy, or not too happy?} \\
\texttt{health} & Health & \detokenize{Would you say your own health, in general, is excellent, good, fair, or poor?} \\
\texttt{life} & Life Outlook & \detokenize{In general, do you find life exciting, pretty routine, or dull?} \\
\texttt{trust} & Social Trust & \detokenize{Generally speaking, would you say that most people can be trusted or that you can't be too careful in dealing with people?} \\
\texttt{satjob} & Job Satisfaction & \detokenize{On the whole, how satisfied are you with the work you do--would you say you are very satisfied, moderately satisfied, a little dissatisfied, or very dissatisfied?} \\
\texttt{satfin} & Income / Satisfaction & \detokenize{We are interested in how people are getting along financially these days. So far as you and your family are concerned, would you say that you are pretty well satisfied with your present financial situation, more or less satisfied, or not satisfied at all?} \\
\texttt{class} & Social Class &  \\
\texttt{parsol} & Parents / Social Mobility & \detokenize{Compared to your parents when they were the age you are now, do you think your own standard of living now is much better, somewhat better, about the same, somewhat worse, or much worse than theirs was?} \\
\texttt{abany} & Abortion & \detokenize{Please tell me whether or not you think it should be possible for a pregnant woman to obtain a legal abortion if the woman wants it for any reason.} \\
\texttt{sexeduc} & Education / Sex & \detokenize{Would you be for or against sex education in the public schools?} \\
\texttt{premarsx} & Sex & \detokenize{If a man and woman have sex relations before marriage, do you think it is always wrong, almost always wrong, wrong only sometimes, or not wrong at all?} \\
\texttt{fear} & Crime / Residence & \detokenize{Is there any area right around here--that is, within a mile--where you would be afraid to walk alone at night?} \\
\texttt{owngun} & Guns & \detokenize{Do you happen to have in your home (IF HOUSE: or garage) any guns or revolvers?} \\
\texttt{god} & Religion & \detokenize{Which statement comes closest to expressing what you believe about God?} \\
\texttt{compuse} & Computers and Internet / Information Society & \detokenize{Do you personally ever use a computer at home, at work, or at some other location?} \\
\texttt{goodlife} & Social Mobility / ISSP Social Inequality & \detokenize{The way things are in America, people like me and my family have a good chance of improving our standard of living -- do you agree or disagree?} \\
\texttt{sexornt} & Sexual Orientation / High Risk Behaviors & \detokenize{Which of the following best describes you?} \\
\texttt{dwelown} & Residence / Dwelling & \detokenize{Do you/Does your family own your (home/apartment), pay rent, or what?} \\
\bottomrule
\end{longtable}
\normalsize

\subsection{NHIS outcomes}
\label{app:nhis-outcomes}
\small
\begin{longtable}{@{}>{\raggedright\arraybackslash}p{0.20\textwidth}>{\raggedright\arraybackslash}p{0.20\textwidth}>{\raggedright\arraybackslash}p{0.54\textwidth}@{}}
\toprule
\textbf{Variable} & \textbf{Category} & \textbf{Original question} \\
\midrule
\endfirsthead
\toprule
\textbf{Variable} & \textbf{Category} & \textbf{Original question} \\
\midrule
\endhead
\texttt{health} & Health Status & \detokenize{Would you say [person's] health in general is excellent, very good, good, fair, or poor?} \\
\texttt{bmicalc} & Physical Health & \detokenize{Body Mass Index calculated from publicly released height and weight variables} \\
\texttt{wldayr} & Health Impact & \detokenize{During the past 12 months, about how many days did illness or injury keep [person] in bed more than half of the day?} \\
\texttt{latotal} & Functional Status & \detokenize{Is [person] limited in any way in any activities because of an impairment or health problem?} \\
\texttt{lany} & Functional Status & \detokenize{Is [person] limited in any way in any activities because of physical, mental, or emotional problems?} \\
\texttt{flany} & Functional Status & \detokenize{Does [person] have any functional limitations?} \\
\texttt{fanylct} & Family Health & \detokenize{Number of family members with any limitation} \\
\texttt{alc1yr} & Substance Use & \detokenize{In [person's] entire life, has [he/she] had at least 12 drinks of any type of alcoholic beverage in any one year?} \\
\texttt{alclife} & Substance Use & \detokenize{In [person's] entire life, has [he/she] had at least 12 drinks of any type of alcoholic beverage?} \\
\texttt{alc5upyr} & Substance Use & \detokenize{In the past year, how many days did [person] have 5 or more drinks of any alcoholic beverage?} \\
\texttt{alcamt} & Substance Use & \detokenize{In the past year, on the days when [person] drank, on the average, how many drinks did [he/she] have?} \\
\texttt{smokev} & Substance Use & \detokenize{Has [person] smoked at least 100 cigarettes in [his/her] entire life?} \\
\texttt{smokagereg} & Substance Use & \detokenize{How old was [person] when [he/she] first started smoking cigarettes fairly regularly?} \\
\texttt{cigdaymo} & Substance Use & \detokenize{During the past 30 days, on how many days did [person] smoke cigarettes?} \\
\texttt{cigsday} & Substance Use & \detokenize{On the average, how many cigarettes does [person] smoke per day?} \\
\texttt{cigsday1} & Substance Use & \detokenize{On the average, how many cigarettes does [person] smoke per day? (Daily smokers)} \\
\texttt{cigsday2} & Substance Use & \detokenize{On the average, how many cigarettes does [person] smoke per day? (Some day smokers)} \\
\texttt{smokestatus2} & Substance Use & \detokenize{Cigarette smoking recode combining current detailed smoking status, former smoking status, and never smoking status} \\
\texttt{smokfreqnow} & Substance Use & \detokenize{Does [person] now smoke cigarettes every day, some days, or not at all?} \\
\texttt{smklsev} & Substance Use & \detokenize{Has [person] ever used smokeless tobacco?} \\
\texttt{smklsfreqnow} & Substance Use & \detokenize{How often does [person] now use smokeless tobacco?} \\
\texttt{ecigev} & Substance Use & \detokenize{Has [person] ever used an e-cigarette, even once or twice?} \\
\texttt{eciged} & Substance Use & \detokenize{Does [person] now use e-cigarettes every day, some days, or not at all?} \\
\texttt{mod10dmin} & Physical Activity & \detokenize{About how long does [person] do these moderate activities each time?} \\
\texttt{mod10fwk} & Physical Activity & \detokenize{In a typical week, on how many days does [person] do moderate activities for at least 10 minutes?} \\
\texttt{vig10dmin} & Physical Activity & \detokenize{About how long does [person] do these vigorous activities each time?} \\
\texttt{vig10fwk} & Physical Activity & \detokenize{In a typical week, on how many days does [person] do vigorous activities for at least 10 minutes?} \\
\texttt{strongfwk} & Physical Activity & \detokenize{In a typical week, on how many days does [person] do physical activities specifically designed to strengthen muscles?} \\
\texttt{hearing} & Sensory Function & \detokenize{Which statement best describes [person's] hearing (without a hearing aid): excellent, good, a little trouble, moderate trouble, a lot of trouble, or deaf?} \\
\texttt{vdifamount} & Sensory Function & \detokenize{How much difficulty does [person] have with vision?} \\
\texttt{hrsleep} & Health Behaviors & \detokenize{On average, how many hours of sleep does [person] get in a 24-hour period?} \\
\texttt{tirefreq3mo} & Health Status & \detokenize{During the past 3 months, how often did [person] feel very tired or exhausted?} \\
\texttt{aeffort} & Mental Health & \detokenize{In the past 30 days, how often did [person] feel that everything was an effort?} \\
\texttt{afeelint1mo} & Mental Health & \detokenize{During the past 30 days, how much did these feelings interfere with [person's] life?} \\
\texttt{ahopeless} & Mental Health & \detokenize{In the past 30 days, how often did [person] feel hopeless?} \\
\texttt{anervous} & Mental Health & \detokenize{In the past 30 days, how often did [person] feel nervous?} \\
\texttt{arestless} & Mental Health & \detokenize{In the past 30 days, how often did [person] feel restless or fidgety?} \\
\texttt{asad} & Mental Health & \detokenize{In the past 30 days, how often did [person] feel sad?} \\
\texttt{aworthless} & Mental Health & \detokenize{In the past 30 days, how often did [person] feel worthless?} \\
\texttt{worfreq} & Mental Health & \detokenize{How often does [person] feel worried, nervous, or anxious?} \\
\texttt{worrx} & Mental Health & \detokenize{Does [person] take medication for worried, nervous, or anxious feelings?} \\
\texttt{worfeelevl} & Mental Health & \detokenize{The last time [person] felt worried, nervous, or anxious, how would [he/she] describe the level of these feelings?} \\
\texttt{depfreq} & Mental Health & \detokenize{How often does [person] feel depressed?} \\
\texttt{deprx} & Mental Health & \detokenize{Does [person] take medication for depression?} \\
\texttt{depfeelevl} & Mental Health & \detokenize{The last time [person] felt depressed, how would [he/she] describe the level of depression?} \\
\texttt{sawment} & Healthcare Utilization & \detokenize{During the past 12 months, did [person] see or talk to a mental health professional such as a psychiatrist, psychologist, psychiatric nurse, or clinical social worker about [his/her] health?} \\
\bottomrule
\end{longtable}
\normalsize

\clearpage
\normalsize
\section{Small-sample hypothesis-testing illustration}
\label{app:small-sample-simulation}

Figure \ref{fig:small-sample} is an analytic design illustration rather than a Monte Carlo simulation over generated outcomes. The calculation considers a standardized outcome with variance equal to one and a true mean difference of \(\mu=0.05\) standard deviations. Survey sample size \(n\) ranges from 50 to 5{,}000, and the two-sided testing level is \(\alpha=0.05\). The survey-only baseline has no auxiliary prediction signal, represented as \(R^2=0\). The three PPI scenarios use \(R^2=0.3\) with \(N_u=10{,}000\), \(R^2=0.6\) with \(N_u=10{,}000\), and \(R^2=0.6\) with \(N_u=100{,}000\).

The calculation uses the large-sample variance approximation for a PPI mean,
\[
\mathrm{Var}(\hat{\mu}_{PPI}) \approx \frac{R^2}{N_u}+\frac{1-R^2}{n}.
\]
The first term is the contribution of the auxiliary prediction mean, and the second term is the survey-estimated residual component. For the survey-only baseline, \(N_u=\infty\) and \(R^2=0\), so the variance reduces to \(1/n\). Effective sample size is defined as
\[
n_{\mathrm{eff}}=\frac{1}{\mathrm{Var}(\hat{\mu}_{PPI})},
\]
because the standardized survey-only mean has variance \(1/n\). The test statistic under the stylized alternative is \(z=\mu/\sqrt{\mathrm{Var}(\hat{\mu}_{PPI})}\), and the reported two-sided p-value is
\[
p=2\{1-\Phi(|z|)\}.
\]
This setup is intentionally simple. It translates a fixed amount of predictive signal and auxiliary-file size into effective sample size and hypothesis-test behavior, holding the target effect size fixed so that Figure \ref{fig:small-sample} isolates the nonlinear consequences of precision gain across survey sample sizes.

\clearpage
\section{Supplementary results}
\label{app:supplementary-results}
\renewcommand{\theHfigure}{supp.\thefigure}
\setcounter{figure}{0}
\renewcommand{\thefigure}{E\arabic{figure}}

\subsection{Pseudo-census simulation}
\label{app:pseudocensus}

\begin{figure}[p]
\centering
\begin{subfigure}{0.95\textwidth}
\centering
\includegraphics[width=0.92\textwidth]{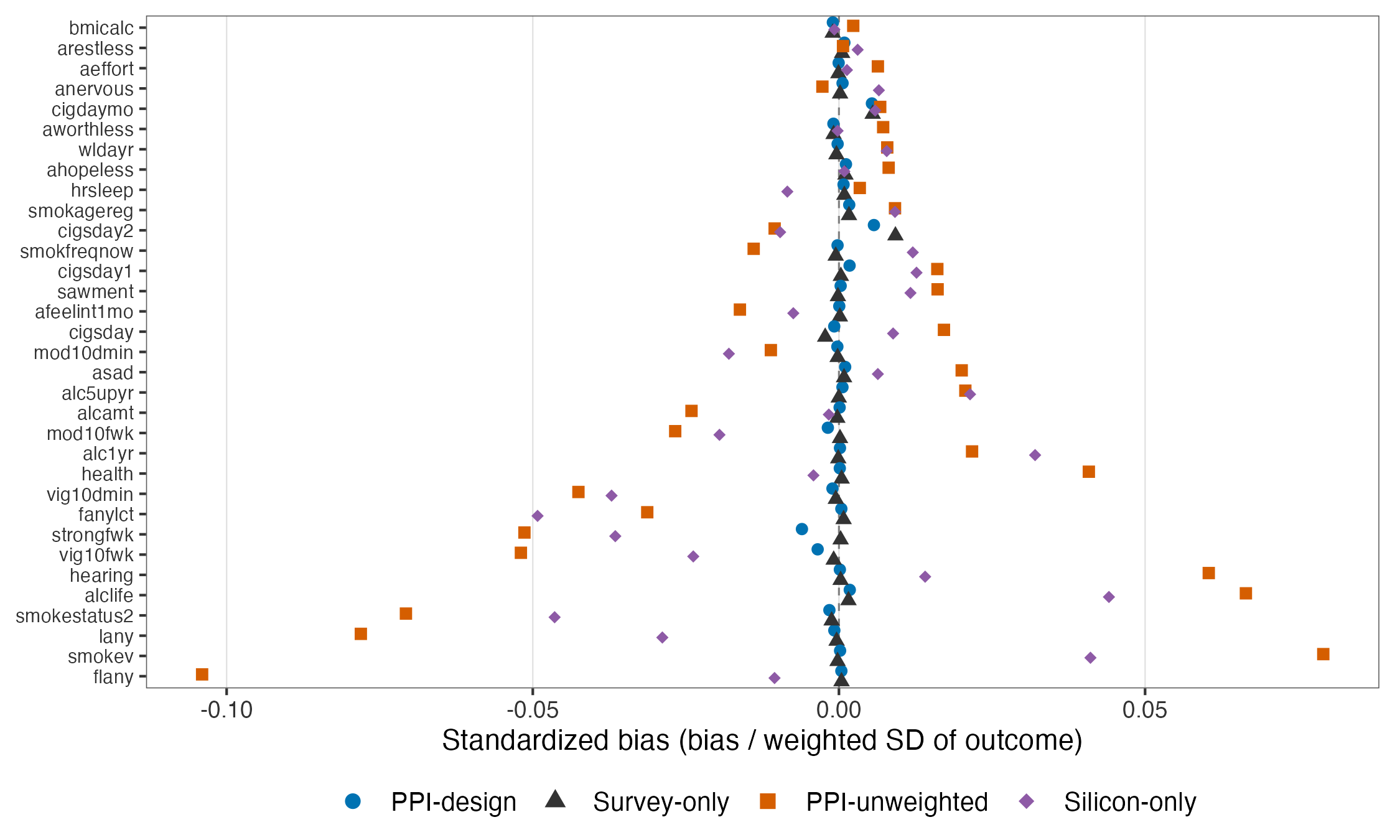}
\caption{Standardized bias}
\end{subfigure}
\vspace{0.5em}
\begin{subfigure}{0.95\textwidth}
\centering
\includegraphics[width=0.92\textwidth]{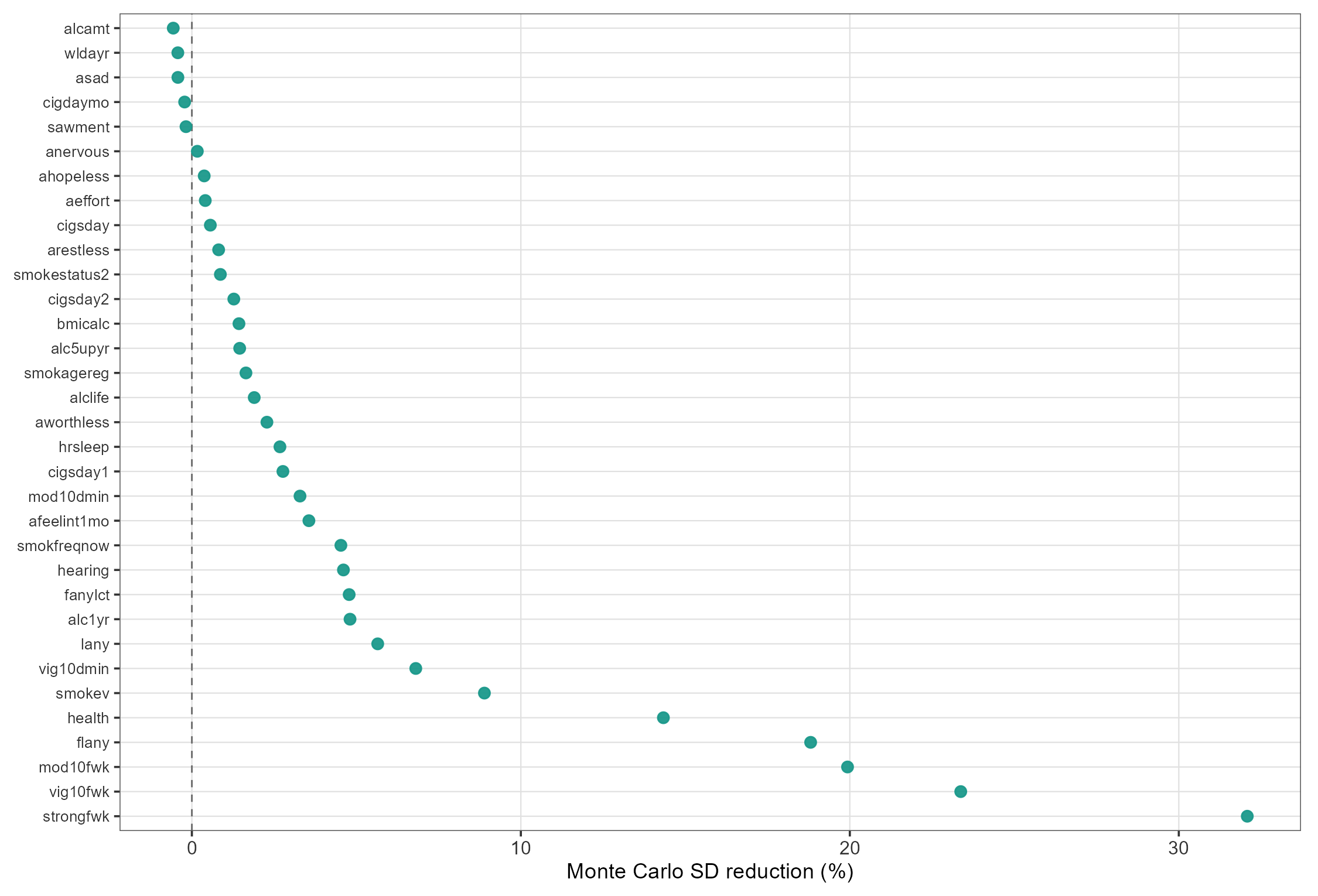}
\caption{Monte Carlo standard-deviation reduction}
\end{subfigure}
\caption{Pseudo-census Monte Carlo using NHIS 2017. The full wave is treated as a finite population; repeated 10 percent subsamples mimic smaller surveys. In panel A, circles denote design-consistent PPI, triangles survey-only estimation, squares unweighted PPI, and diamonds prediction-only estimation. The dashed line indicates zero bias. Panel B reports percentage reductions in Monte Carlo standard deviation relative to survey-only estimation.}
\label{fig:pseudocensus}
\end{figure}

This supplementary analysis directly evaluates bias and repeated-sampling variability against a known finite-population target, complementing the point-estimate comparisons and expected efficiency gains in the survey applications. Figure \ref{fig:pseudocensus} uses the NHIS 2017 validation wave as a pseudo-census and compares four estimators across repeated 10 percent subsamples. Here, predictions come from historical supervised learning, a learner trained on all data in NHIS prior to 2017. Panel A reports standardized bias. Survey-only and design-consistent PPI estimates are close to zero. Across outcomes, the largest absolute standardized bias is about 0.009 for survey-only estimation and 0.006 for design-consistent PPI, compared with about 0.049 for silicon-only estimation and 0.104 for conventional unweighted PPI. The results suggest that more predicted observations do not remove model bias, and the residual correction must respect the survey design.

Panel B adds the precision result that compares the Monte Carlo standard deviation of PPI estimates and survey-only estimates. Design-consistent PPI reduces Monte Carlo sampling variability by about 32 percent for days of muscle-strengthening activity, 23 percent for days of vigorous activity, and 20 percent for days of moderate activity. The gain is much smaller for weakly predicted outcomes. The simulation demonstrates the benefit of applying PPI. Compared with survey-only estimates, design-consistent PPI estimates are similarly centered but more precise. Prediction-only estimates can be biased and therefore should not be used to infer the population directly.

\clearpage
\subsection{Prediction-only and corrected point estimates}
\label{app:estimate-anchoring}

\begin{figure}[H]
\centering
\includegraphics[width=0.94\textwidth]{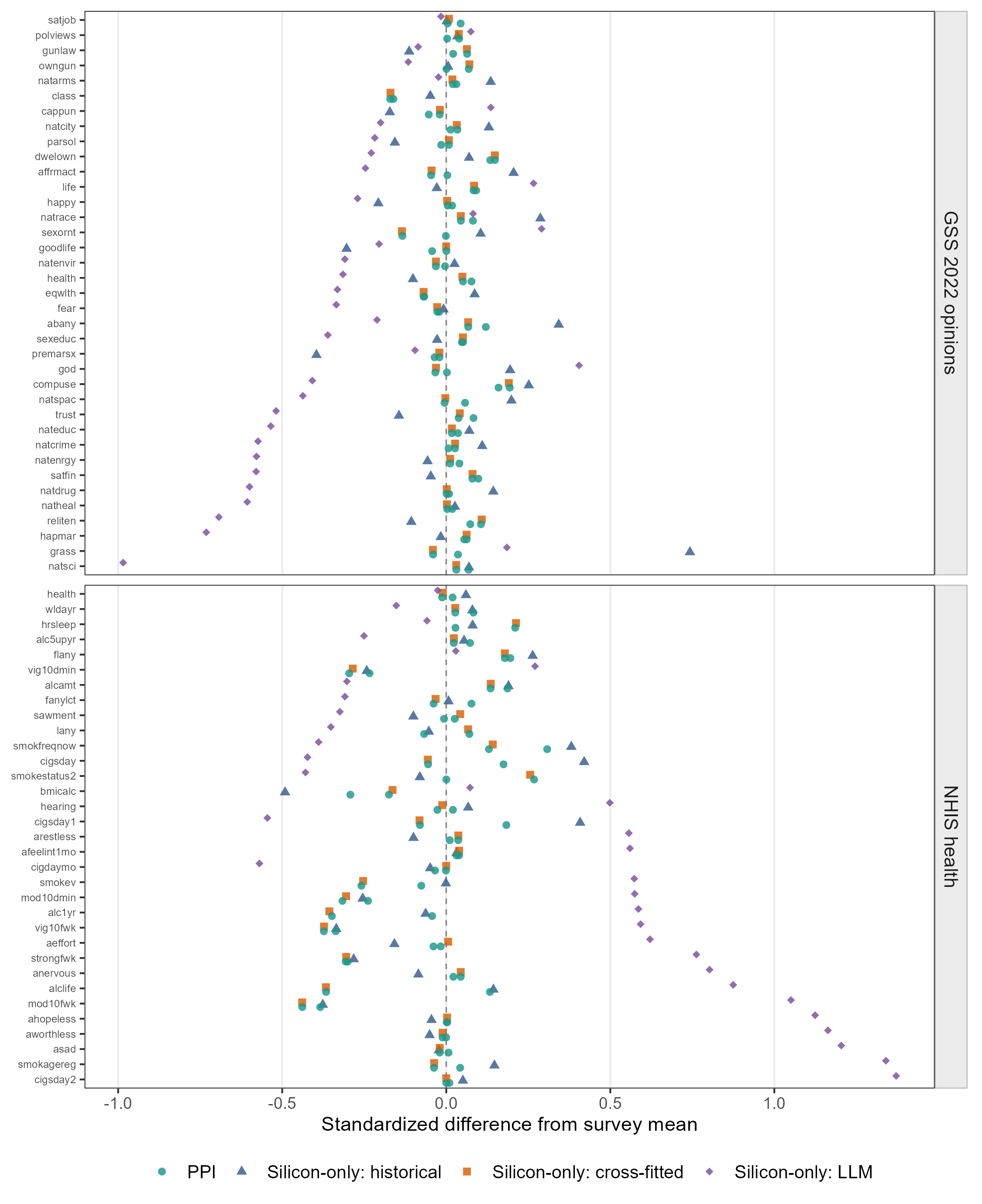}
\caption{Prediction-only and residual-corrected outcome means relative to observed survey means. The horizontal axis reports the estimate minus the survey mean, divided by the survey standard deviation of the outcome. All means in this descriptive comparison are unweighted and use the corresponding valid cases. Historical and cross-fitted supervised prediction means are computed over ACS; their corrected PPI estimates appear as green circles. The LLM series reports prediction-only means for survey respondents with valid LLM predictions. The dashed line marks agreement with the observed survey mean.}
\label{fig:estimate-anchoring}
\end{figure}

Figure \ref{fig:estimate-anchoring} shows how predictions change estimated outcome levels in the GSS and NHIS applications. Figure \ref{fig:pseudocensus} evaluates bias and sampling variability against a known population truth, while Table \ref{tab:summary-gains} and Figures \ref{fig:nhis-eff} and \ref{fig:gss-eff} report precision gains. Comparing prediction-only and corrected PPI estimates with observed survey means illustrates the difference between replacing survey responses and augmenting them with predictions. Each point reports the difference from the survey mean in outcome standard deviations. Positive values indicate higher estimates; negative values indicate lower estimates. Zero marks agreement with the observed survey mean, rather than a known population truth.

The corrected PPI estimates differ relatively little from the survey means across outcomes, with a median absolute standardized difference of about 0.04 and a 90th percentile of about 0.24. Prediction-only estimates sometimes differ substantially more: the largest absolute differences for the LLM series are about 1.37 outcome standard deviations in NHIS and 0.98 in GSS. PPI does not require the corrected estimate to equal the survey mean. Its remaining difference reflects the difference between the auxiliary-file and survey prediction means after the survey's average prediction error has been removed.

\end{document}